\documentclass[10pt,letterpaper]{article}
\usepackage[top=0.75in,left=0.65in,right=0.65in,footskip=0.85in]{geometry}
\pdfoutput=1

\newcommand{\e}{\text{e}}
\newcommand{\dd}{\text{d}}

\usepackage{amsmath,amssymb,amsthm}

\usepackage{bbm}
\usepackage{geometry}

\usepackage{amsmath,amssymb}

\usepackage{textcomp,marvosym}

\usepackage{cite}

\usepackage{nameref,hyperref}

\usepackage{wrapfig}

\usepackage[right]{lineno}

\usepackage{ulem}

\usepackage[table]{xcolor}

\usepackage{array}

\newcolumntype{+}{!{\vrule width 2pt}}

\newlength\savedwidth

\usepackage[aboveskip=1pt,labelfont=bf,labelsep=period,justification=raggedright,singlelinecheck=off]{caption}

\makeatletter
\renewcommand{\@biblabel}[1]{\quad#1.}
\makeatother

\usepackage{lastpage,fancyhdr,graphicx}
\usepackage{epstopdf}
\usepackage{algorithmic,algorithm}

\fancyheadoffset[L]{2.25in}
\fancyfootoffset[L]{2.25in}
\begin{document}
%\twocolumn

\title{Networks of Necessity: Simulating COVID-19 Mitigation Strategies for Disabled People and Their Caregivers}
\author{Thomas E. Valles \thanks{Department of Mathematics, University of California, Los Angeles} \and Hannah Shoenhard \thanks{Department of Cell and Developmental Biology, University of Pennsylvania} \and Joseph Zinski \footnotemark[2] \and Sarah Trick \thanks{Assistant Editor at tvo.org (TVOntario)} \and Mason A. Porter \footnotemark[1] \thanks{Santa Fe Institute} \and Michael R. Lindstrom \footnotemark[1] \thanks{corresponding author: mikel@math.ucla.edu}}
\date{\today}

\maketitle

%%%%%%

% ABSTRACT

\begin{abstract}

A major strategy to prevent the spread of COVID-19 is the limiting of in-person contacts. However, limiting contacts is impractical or impossible for the many disabled people who do not live in care facilities, but still require caregivers to assist them with activities of daily living. We seek to determine which interventions can prevent infections among disabled people and their caregivers. To accomplish this, we simulate COVID-19 transmission with a compartmental model that includes susceptible, exposed, asymptomatic, symptomatically ill, hospitalized, and removed/recovered individuals. The networks on which we simulate disease spread incorporate heterogeneity in the risks of different types of interactions, time-dependent lockdown and reopening measures, and interaction distributions for four different groups (caregivers, disabled people, essential workers, and the general population). Among these groups, we find that the probability of becoming infected is largest for caregivers and second largest for disabled people. Consistent with this finding, our analysis of network structure illustrates that caregivers have the largest modal eigenvector centrality among the four groups. We find that two interventions --- contact-limiting by all groups and mask-wearing by disabled people and caregivers --- most reduce the cases among disabled people and caregivers. We also test which group of people spreads COVID-19 most readily by seeding infections in a subset of each group and comparing the total number of infections as the disease spreads. We find that caregivers are the most potent spreaders of COVID-19, particularly to other caregivers and to disabled people. We test where to use limited vaccine doses most effectively and find that (1) vaccinating caregivers better protects disabled people than vaccinating the general population or essential workers and that (2) {vaccinating caregivers protects disabled people about as much as vaccinating disabled people themselves}. Our results highlight the potential effectiveness of mask-wearing, contact-limiting throughout society, and strategic vaccination for limiting the exposure of disabled people and their caregivers to COVID-19.

\end{abstract}

{\bf keywords:} COVID-19, disabilities, networks, contagions, parameter estimation, vaccination

%%%%%%

% AUTHOR SUMMARY

\section*{Author Summary}

Disabled people who need help with daily life tasks, such as dressing or bathing, have frequent close contacts with caregivers. This prevents disabled people and their caregivers from physically distancing from one another, and it also significantly increases the risk of both groups to contract COVID-19. 
How can society help disabled people and caregivers avoid infections? To answer this question, we simulate infections on networks that we model based on a city of about one million people. We find that one good strategy is for both disabled people and their caregivers to use masks when they are together. 
We also find that if only disabled people limit their contacts while other people continue their lives normally, disabled people are not effectively protected. However, it helps disabled people substantially if the general population also limits their contacts. We also study which vaccination strategies can most efficiently protect disabled people. Our simulations suggest that vaccinating caregivers against COVID-19 protects the disabled subpopulation about equally effectively as vaccinating a similar number of disabled people. Our findings highlight both behavioral measures and vaccination strategies that society can take to protect disabled people and caregivers from COVID-19.

%%%%%%

%INTRO

\section{Introduction}

The coronavirus disease 2019 (COVID-19) pandemic, which is caused by the severe acute respiratory syndrome coronavirus 2 (SARS-CoV-2) virus, has revealed major societal vulnerabilities in pandemic preparation and management \cite{Maxmen2021}. Existing social disparities and structural factors have led to a particularly adverse situation for the spread of COVID-19 in vulnerable groups. Therefore, it is crucial to examine how to mitigate its spread in these vulnerable groups \cite{NatureEditorial2020} both to address these difficulties in the current pandemic and to prepare for future pandemics \cite{Zelner2021}. The effectiveness of society-wide behavioral interventions in mitigating viral spread in the general population is now well-documented \cite{Flaxman2020, Dehning2020, Alfano2020, chu2020physical, VanDyke2020}. However, the effectiveness of these non-pharmaceutical interventions (NPIs) has not been assessed in certain vulnerable groups. One such group is disabled people, who may choose to live in a group-care setting (such as a nursing home) or live independently with some caregiver support. It has been speculated that the latter arrangement increases the risk of disabled people to exposure to infections \cite{Shakespeare2021}. However, to the best of our knowledge, this situation has not been studied using epidemiological modeling. Vaccinations have also been extraordinarily effective at mitigating COVID-19; they have decreased case numbers and case rates, onset of symptomatic disease, hospitalizations, and mortality numbers and rates \cite{Chodick2021,Thompson2021, LACounty2021, PfizerClinical2021, ModernaClinical2021,JJClinical2021,AZClinical2021}. However, strategies for how to most efficiently use vaccines to protect independently housed disabled people have not yet been evaluated. In the present paper, we study a compartmental model of COVID-19 spread on a network to examine the effectiveness of several non-pharmaceutical interventions (NPIs) and vaccination strategies to prevent the spread of COVID-19 among independently housed disabled people and their caregivers. 

People with disabilities who require assistance with activities of daily living (ADLs) may live in a long-term care facility or independently with some form of caregiving support \cite{Young2017, Gorges2019}. Although extensive epidemiological and modeling studies have identified risk factors and mitigation strategies for COVID-19 outbreaks in long-term care facilities \cite{HeNurse2020, KimNurse2020, AbramsNurse2020, LiNurse2020, ChenNurse2020, GorgesNurse2020}, there have not been similar studies of independently housed disabled people and their caregivers. Caregivers are often indispensable for the health and independence of disabled people because they assist with activities such as bathing, dressing, and using the bathroom. However, in a pandemic, public-health concerns dictate that it is important to minimize in-person contacts. Disabled people and their caregivers thus face an urgent question: How can they continue to interact while minimizing the risk of COVID-19 transmission?

\begin{figure}
        \centering
        \includegraphics[width=4in]{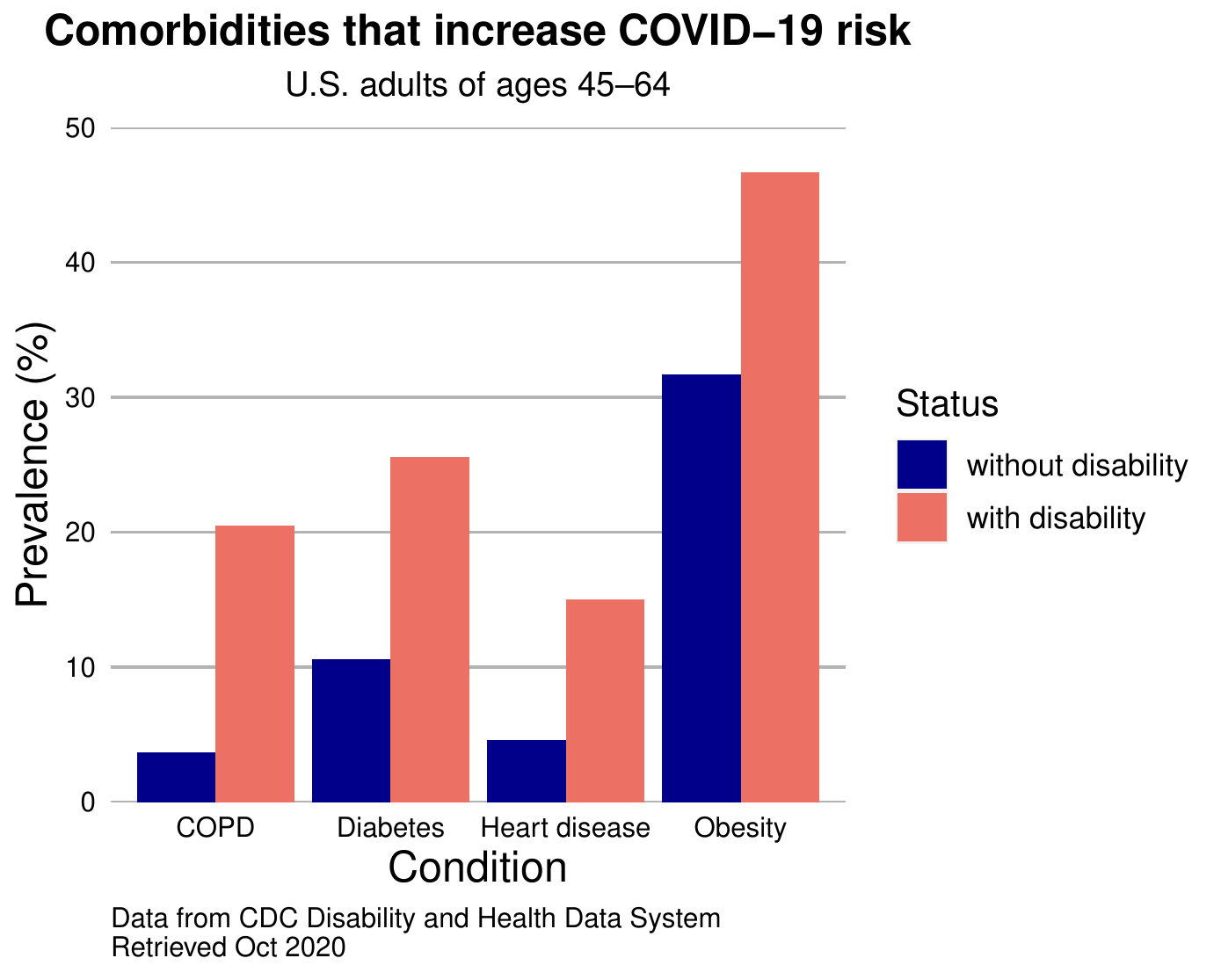}
        \caption{Rates of comorbidities that predispose individuals (of ages 45--64) to severe cases of COVID-19 among adults in the United States without (blue) and with (red) disabilities.}
        \label{fig:CoMo}
    \end{figure}

This question is especially urgent because of the high prevalence of risk factors for severe COVID-19 in the disabled population. These risk factors, for which we give statistics for adults of ages 45--64 in the United States (see Fig.~\ref{fig:CoMo}) \cite{CDCWarns2020, DisabilityStats2020}, include obesity (about 46.7\% of adults with a disability have a body mass index (BMI) that indicates obesity, compared with about 31.7\% of adults without a disability), heart disease (15.0\% of adults with a disability and 4.6\% of adults without one), Chronic Obstructive Pulmonary Disease (COPD) (20.5\% of adults with a disability and 3.7\% of adults without one), and diabetes (25.6\% of adults with a disability and 10.6\% of adults without one). Additionally, whatever factor initially causes a person's disability can also complicate medical management of their case if they contract COVID-19. Furthermore, isolating while ill can be impossible for disabled people because they rely on caregivers to assist them with essential daily tasks. This can make disabled people more prone to spread COVID-19 to caregivers if they contract it. Consequently, preventing COVID-19 infection among the disabled population and caregivers should be a high priority.

Caregivers also experience high risk of exposure to and death from COVID-19. Caregiving workers are disproportionately likely to be women, immigrants, and people of color. The median wage for in-home caregivers is \$12.12 per hour, and their median annual earnings are \$17,200 (which is below the U.S. federal poverty guideline for a two-person household) \cite{PHI2020}. Experiencing poverty or being Black or Latinx independently increase the risk because of systemic disadvantages in accessing healthcare \cite{Adhikari2020, Egede2020, Kim2020}. Furthermore, the COVID-19 pandemic has brought immense challenges to the caregiving workforce, including frequent lack of personal protective equipment (PPE), pandemic-specific training, paid time off, and childcare \cite{PHI2020}. Finally, much caregiving work is impossible without close physical contact, which elevates caregivers' risk of occupational exposure. In summary, caregivers often belong to groups that are at higher risk both of COVID-19 exposure and of more severe illness from it. 

According to a 2018 report \cite{Disabled2018}, approximately 26\% of U.S. adults (including about 41\% of those who are 65 or older) have some form of disability. In 2016, Lauer and Houtenville \cite{lauer20182017} reported that 7.3\% of the American population have a cognitive or physical disability that causes difficulty in dressing, bathing, or getting around inside the home (but we acknowledge the large uncertainty in this estimate). At least 2.4 million people in the U.S. (i.e., approximately 0.7\% of the population) are employed as home-care workers, but this is likely an underestimate because of the difficulty of accurate statistical collection \cite{PHI2020}. An intense time commitment and irregular hours are necessary for care, so many disabled people hire multiple caregivers and many caregivers work for multiple disabled people \cite{NJDisability2020}. Therefore, there is significant potential for the spread of COVID-19 among and between these two vulnerable populations, making it a high priority to identify effective methods to reduce COVID-19 spread among disabled people and caregivers without compromising care.

To mitigate disease spread during a pandemic, governments may choose to implement society-wide shutdown orders, mask mandates, and/or physical-distancing guidelines. However, governments in some regions have been reluctant to issue such orders, and populations may not fully comply with them. This raises the issue of what disabled people and caregivers can do to protect themselves both with and without society-wide pandemic-mitigation efforts. With this in mind, we test how effectively mask-wearing (i.e., using PPE), limiting the number of caregiver contacts, and limiting contacts among disabled people prevent COVID-19 infections when the general population either maintains their normal contact levels or limits them. To the best of our knowledge, this is the first time that mathematical modeling has been used to evaluate these issues for COVID-19 infections.

Multiple COVID-19 vaccines are now widely available in some countries, but vaccine supplies remain scarce in other countries. As of late August 2021, only 1.6\% of people in low-income countries have received at least one dose of any COVID-19 vaccine \cite{owidcoronavirus}. Furthermore, other pandemics may well emerge in the future.
Consequently, it is valuable to evaluate how to most effectively allocate a small number of vaccine doses to protect vulnerable groups, such as disabled people. Specifically, we investigate whether vaccinating disabled people or caregivers is more effective than other vaccination strategies for reducing the total number of cases in these two vulnerable groups.
  
In this paper, we simulate COVID-19 spread on model networks that represent a city. We base the parameter values in these networks on Ottawa, Canada. Our stochastic model of disease spread takes into account several disease states (i.e., ``compartments''), different occupation types in a population, the heterogeneity of the risk across different interactions, and time-dependent lockdown measures. Our disease-spread model, which we explain in Section \ref{sec:model}, allows us to quantitatively study our various questions under our set of assumptions. Using both computations of network structure and simulations of the spread of a disease on our networks, we find that disabled people and caregivers are both substantially more vulnerable to COVID-19 infection than the general population because of their large network centralities. We test the effectiveness of several NPIs --- including limiting the number of social contacts, wearing masks, and limiting the number of caregivers that a given disabled person sees --- at preventing COVID-19 spread among disabled people and their caregivers. By selectively seeding infections or blocking infections (via a simulated vaccine) in certain groups, we identify caregivers as major drivers of COVID-19 spread --- especially among disabled people and caregivers --- and suggest that this group should be prioritized in vaccination campaigns. 

Our paper proceeds as follows. We present our stochastic model of the spread of COVID-19 in Section \ref{sec:model}, our results and a series of case studies in Section \ref{sec:results}, and our conclusions and further discussion in Section \ref{sec:discussion}. We describe the details of our model in Sections \ref{sec:estimate} and \ref{sec:simulate} of our Supporting Information.

%%%%%%

% MODEL

\section{A Stochastic Model of the Spread of COVID-19 Infections}
\label{sec:model}

We start by giving a rough idea of our stochastic model of the spread of COVID-19, and we then discuss further details in Section \ref{sec:modelSpecific}. Readers who are interested predominantly in the essence of our model can safely skip Section \ref{sec:modelSpecific}. We give a comprehensive list of our assumptions in Section \ref{sec:modelAssumptions}. Readers who wish to use our code can find it \href{https://3k1m@bitbucket.org/3k1m/covid19-disabledcaregiverstudy.git}{at a Bitbucket repository}. We previously wrote a white paper about this topic\cite{ourWhitePaper}; the present manuscript gives the full details of our study.

%%%%

\subsection{A Brief Overview of Our Model}\label{sec:modelHigh}

Numerous researchers have used mathematical approaches to examine the spread of COVID-19 \cite{estrada2020,arino2021}. Such efforts have used a variety of frameworks, including compartmental models \cite{zaplotnik2020,sameni2020mathematical}, self-exciting point processes \cite{Browning2020,escobar2020hawkes} (which one can also relate to some compartmental models \cite{bertozzi2020}), and agent-based models \cite{hoertel2020agent}. Many of these models incorporate network structure to examine how social contacts affect disease spread.
Some models have incorporated age stratification \cite{Arenas2020}, how mobility and other data can forecast the spread of COVID-19 \cite{barbosa2020mobility,Kraemer493,bertozzi2020challenges}, and/or the structure of travel networks \cite{lai2020}. In the present paper, we use an agent-based approach to study COVID-19 within a single city. Our approach involves simulating a stochastic process on time-dependent networks \cite{holme2012temporal,dakiche2019tracking}. One of the features of our model is that different segments of the population have different degree distributions, with mixing between these different segments. To examine networks with these features, we use generalizations of configuration models \cite{melnik2014dynamics,miller2013incorporating}. 

%%%%%

In our model population, we consider three types of interactions between individuals, six disease states, and four distinct groups (i.e., subpopulations). We encode interactions using a network, and all interactions between different individuals involve exactly two people.
We suppose that {\it strong} interactions describe interactions at home within family units (or, more generally, within ``household units''); {\it weak} interactions describe social interactions and interactions that take place at work, at a grocery store, and so on; and {\it caregiving} interactions describe interactions between caregivers and the disabled people for whom they care. We model each of these interactions with a different baseline level of risk of disease transmission. Weak interactions have the lowest baseline risk level, strong interactions have the next-lowest baseline risk level, and caregiving interactions have the highest baseline risk level.

We use a compartmental model of disease dynamics \cite{Brauer2019}, which we study on contact networks \cite{kiss2017,PastorReview2015}. We assume that our population (e.g., of a single city, like Ottawa) is closed and that each individual is in exactly one disease state (i.e., ``compartment''). Our model includes {\it susceptible (S)} individuals, who can contract COVID-19; {\it exposed (E)} individuals, who have the disease but are not yet infectious or symptomatic; {\it asymptomatic (A)} individuals, who do not have symptoms but can spread the disease; {\it ill (I)} individuals, who are symptomatically ill and contagious; {\it hospitalized (H)} individuals, who are currently in a hospital; and {\it removed (R)} individuals, who are either no longer infectious or have died from the disease. The A compartment includes prodromal infections, asymptomatic individuals, and mildly symptomatic individuals; in all of these situations, an individual has been infected, but we assume that they are not aware of it. Our model does not incorporate loss of immunity or births, and we classify both ``recovered'' and removed individuals as part of the R compartment. In our study, an individual has been infected if they are no longer in the S compartment. Therefore, cumulative infections include every individual that is currently in the E, A, I, H, and R compartments.

We divide our model city's population into the following subpopulations:
\begin{itemize}
\item {\it caregivers}, who provide care to disabled people;
\item {\it disabled people}, who receive care;
\item {\it essential workers}, whose occupations prevent them from limiting contacts during lockdowns and similar policies, but who are not already included in the caregiving population; and
\item the {\it general population}, which is everyone else.
\end{itemize}
The individuals in the disabled subpopulation have two types of caregivers: {\it weak} caregivers, who are professional caregivers whose connections are likely to break if either individual in an interaction becomes symptomatic; and {\it strong} caregivers, whose caregiving relationship persists even if the individuals in it are symptomatic (and as long as neither individual is hospitalized). We consider these two types of caregivers to account for family members or close friends who always provide some care to a disabled person.
Although our model includes a hospitalized compartment, we do not model doctors, nurses, custodial services, or other hospital staff who are involved in caring for COVID-19 patients. The caregivers in our model population refer strictly to individuals who provide supportive assistance to members of the disabled community in their homes. We also do not model skilled care facilities, such as nursing homes.

When an individual is symptomatic, we assume that they distance themselves (through so-called ``physical distancing'' or ``social distancing'') from society with a fixed probability $b \in [0,1]$. The probability can be less than $1$ to account for a variety of situations, such as people who feel financial pressure to work anyway \cite{presenteeismTasmania}, people who have symptoms that are so mild that they are unaware of them, and people who ignore common decency. In our model, distancing by an individual who becomes ill means that they temporarily cut off their weak contacts or weak caregiver--disabled relationships and only maintain contacts within their household unit and possibly strong caregiver--disabled relationships until they recover. If an individual becomes hospitalized, these stronger contacts also break.

We seek to understand how COVID-19 spreads in these different groups over time and how different mitigation strategies, such as contact-limiting and mask-wearing, affect the outcomes. Consequently, we allow the distributions of the number of contacts to change with time and adjust the disease transmission probability to reflect the presence of masks.

\begin{figure}
\begin{center}
\includegraphics[width=6.in]{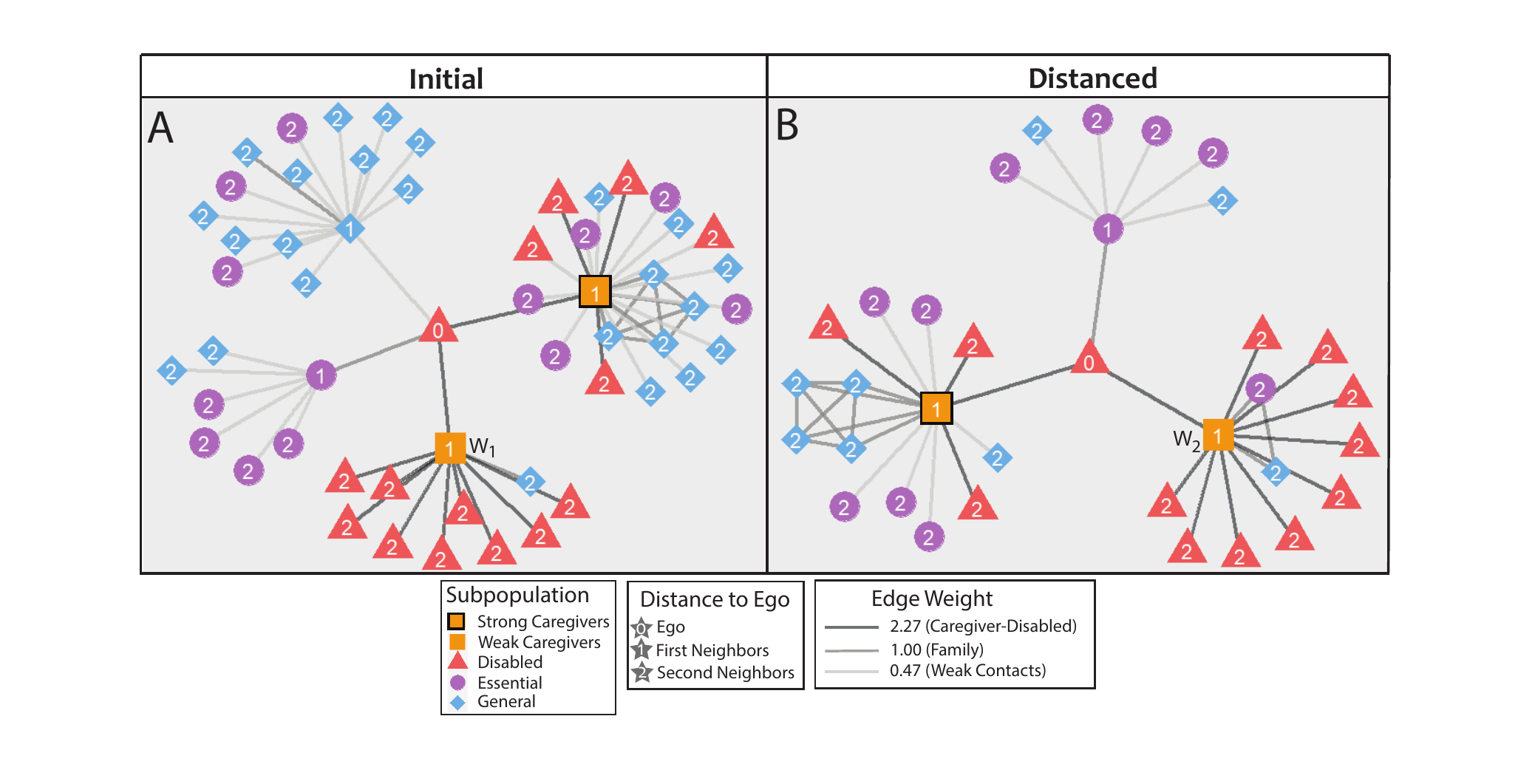}
\smallskip
\caption{An egocentric network (i.e., ego network) of an example disabled person on (A) day 43 (before the start of contact-limiting) and (B) day 45 during contact-limiting). The two ego networks encode contacts for the same disabled person. The label `W\textsubscript{1}' denotes the weak caregiver on day 43 and the label `W\textsubscript{2}' denotes the weak caregiver on day 45. (In this example, W\textsubscript{1} and W\textsubscript{2} are different caregivers. We illustrate the different groups in our model city (colors), interaction strengths between individuals (line thicknesses), and {distances (numbers) from the ego}. The edge weights are relative to the strong-contact weight of $1$.} 
\label{fig:ego}
\end{center}
\end{figure}

We tune our baseline model to describe the city of Ottawa from its first reported case on 10 February 2020 \cite{ottawa_case_counts} through its closure of non-essential businesses on 24 March 2020 \cite{global_close} (the closure order occurred on 23 March) and then to understand how its ``Phase 1'' reopening on 6 July 2020 \cite{ottawa_reopen} affected disease spreading. In Fig.~\ref{fig:ego}, we illustrate an egocentric network (i.e., ``ego network'') \cite{baek2020} that is centered at a single disabled person in the population before and after closure.

%%%%

\subsection{Specific Details of Our Model} \label{sec:modelSpecific}

We now give a detailed description of our model. One of the key elements of the networks on which a disease spreads is their ability to describe the numbers and distributions of the contacts of different types of individuals. We do this by constructing networks using a generalization of configuration-model networks. See \cite{fosdick2018} for a review of configuration models.

To each node (i.e., individual) in a network, we assign a group (disabled, caregiver, essential worker, or general) and then assign both weak contacts and strong contacts. Additionally, we assign caregiver nodes to each disabled node and assign disabled nodes to each caregiver node. No individual is both a strong contact and a weak contact to the same person; no individual is both a caregiver for a disabled person and an ordinary weak contact for that disabled person, and so on. We anticipate a large variance in the number of weak contacts, with some people having many more contacts than others \cite{newman2018networks}, so we assign each individual a number of weak contacts from an approximate truncated power-law distribution (see Section \ref{sec:simulate}). Because strong contacts represent household units, we assign each individual a number of strong contacts from an empirical distribution that we construct using census data of household sizes in Canada \cite{census}. To model the pools of caregivers that are available to disabled people, we assume that each disabled node has a fixed number of weak caregivers (this pool does not change over time) and that this fixed number is the same for all disabled nodes. We were unable to find reliable data about the sizes of these pools, so we base the values of these quantities on educated guesses that are consistent with the lived experience of the disabled authors of the present paper. We also assign one strong caregiver to each disabled node. The contact structure in one of our networks can change over time. For example, weak contacts can break if a lockdown starts, both weak and strong contacts break when an individual is hospitalized, and so on. Each day, we choose one member of a disabled individual's caregiver pool uniformly at random to potentially provide care to them. (It is only potential care because that caregiver may have broken contact due to illness.) Each day, the disabled individual also receives care from a single strong caregiver, if possible. (This occurs as long as that contact has not been broken due to hospitalization.) In each time step, which consists of one day, the disease state (i.e., compartment) of an individual can change. From one day to the next, we compute the transition probability from susceptible to exposed using Eqs.~\eqref{eq:notget} and \eqref{eq:get} using each susceptible individual's disease state at the start of the day. On each day, we determine transitions between different disease states by generating exponential random variables for transition times. When a generated transition time occurs within a $1$-day window, an individual changes compartments. If two different transitions are possible and both exponential random variables are less than $1$ day, then we use the state transition that corresponds to the shorter transition time. Individuals who break their contacts because of illness do so immediately upon transitioning to a new compartment. Any network restructuring occurs at the start of a day (i.e., before we calculate exposure risks).

In the Supporting Information, we give the day-to-day evolution in Algorithms \ref{alg:overall}, \ref{alg:infect}, and \ref{alg:advance} and the network-construction process in Algorithms \ref{alg:unit}, \ref{alg:weak}, \ref{alg:strong}, and \ref{alg:disabled}. We host our code at a \href{https://3k1m@bitbucket.org/3k1m/covid19-disabledcaregiverstudy.git}{Bitbucket repository}.

When we construct our networks, we use three families of discrete probability distributions.
\begin{itemize}
\item The distribution $\mathcal{P}(a_-,a_+,\mu)$ is an approximate truncated power-law distribution. If $N \sim \mathcal{P}(a_-,a_+,\mu)$, then $N$ takes integer values in $\{a_-, a_-+1, \ldots, a_+ \}$. For large $n$, we have that $\Pr(N=n) = O(1/n^p)$, where we choose $p$ so that $\mathbb{E}(N) = \mu$. See our Supporting Information for full details.
\item The distribution $\mathcal{E}(p_0, p_1, p_2, \ldots, p_k)$ is a discrete distribution. If $N \sim \mathcal{E}(p_0, p_1, p_2, \ldots, p_k)$, then $\Pr(N = n) = p_n$ when $n \in \{0, \ldots, k\}$ and $\Pr(N = n) = 0$ otherwise.
\item The distribution $\mathcal{F}(k)$ is a deterministic distribution that has only one attainable value. If $N \sim \mathcal{F}(k)$, then $\Pr(N = n) = \delta_{n,k}$, where $\delta$ denotes the Kronecker delta (which equals $1$ if the subscripts are equal and $0$ if they are not).
\end{itemize}

Our model has three key dates: the {\it first recorded case},
which we set to be day 1 (i.e., 10 February 2020), as we use day $0$ for initial conditions to produce a seed case of the disease (such cases are in the asymptmatic compartment); {\it lockdown} (i.e., 24 March 2020), which is when contact-limiting begins and some individuals wear masks; and {\it reopening} (i.e., 6 July 2020), which is when the city begins to reopen. For mask-wearing, we focus on four situations: 
{
\begin{itemize}
\item None (N): nobody wears a mask;
\item Disabled people and caregivers wear masks (D+C): disabled people and caregivers both wear masks when interacting with each other, but nobody else wears a mask; 
\item Disabled people, caregivers, and essential workers wear masks (D+C+E): all of the mask-wearing in the (D+C) scenario occurs, and we also assume that both individuals in an interaction wear a mask whenever there is a weak interaction with an essential worker (to model interactions in places like grocery stores, banks, and routine doctor visits); and
\item All weak contacts wear masks (All*): the same individuals under the same conditions as in (D+C+E) wear masks, but we also assume that both individuals wear a mask in any interaction between weak contacts.
\end{itemize}}

To model essential workers, we assume (except when there are symptoms of illness) that weak contacts with essential workers are not broken. Therefore, during a lockdown, essential workers continue to have a large number of weak contacts on average. We similarly characterize the caregiver subpopulation; they retain their interactions with their associated disabled nodes. We assume that an individual's weak contacts during lockdown are a subset of their weak contacts from before a lockdown. Upon a reopening, each individual is assigned a new number of weak contacts. They retain the weak contacts that they had during a lockdown, but they can also gain new weak contacts that they did not possess before the lockdown if their new number of weak contacts is larger than their number of contacts immediately prior to reopening. For example, if an individual has 3 weak contacts during lockdown and are assigned 7 weak contacts after reopening, then they need 4 weak contacts. These 4 new weak contacts can be different from that individual's weak contacts before the lockdown. We do this to account for situations (such as business closures or job loss) that cause individuals to visit different stores or workplaces after a city reopens.

\begin{figure}
\begin{center}
\includegraphics[width=3.5in]{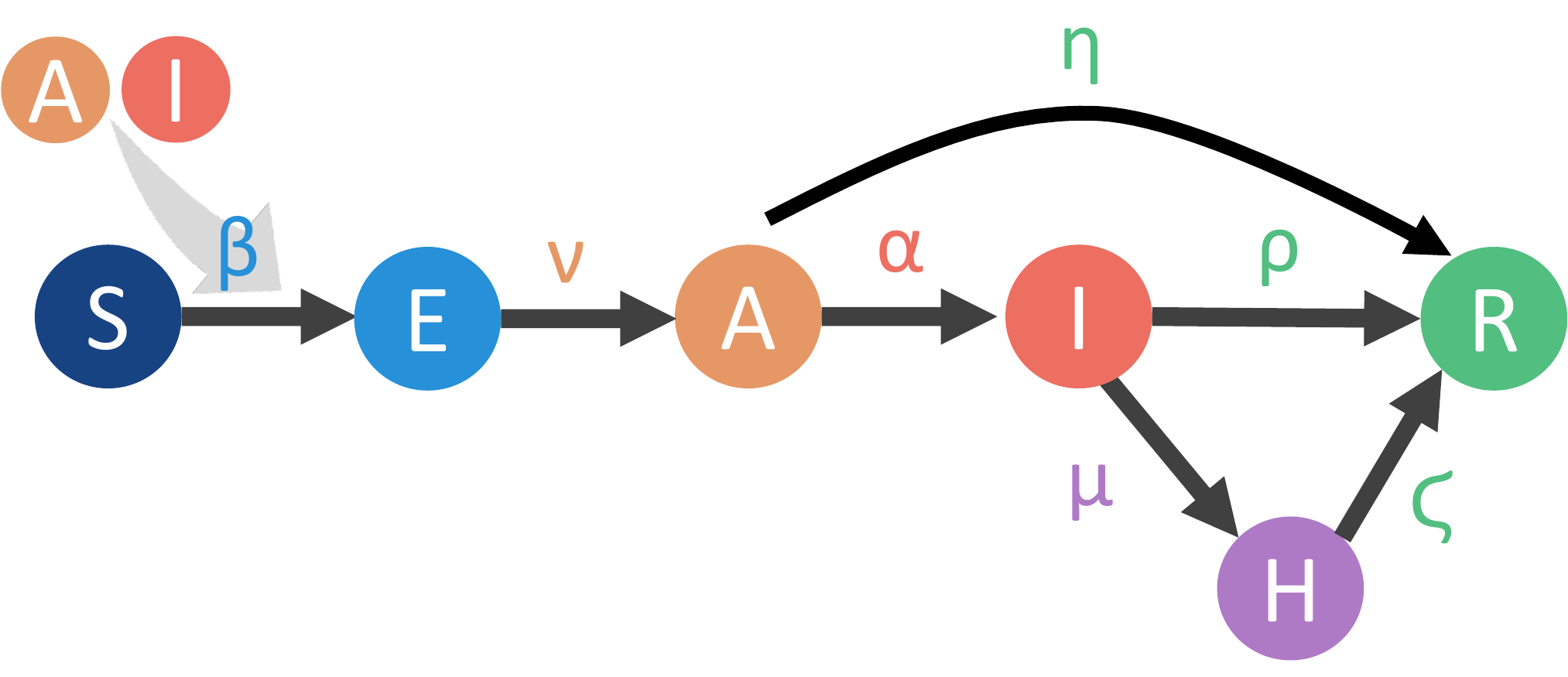}
\caption{Schematic illustration of our compartmental model of disease transmission. Susceptible individuals (S), by being exposed to asymptomatic (A) or symptomatically ill (I) individuals, can become exposed (E) with a baseline transmission probability $\beta$. One can reduce the risk of an interaction through the NPI of mask-wearing; this multiplies the risk by the factor $m^{1/2}$ (if only one individual in the interaction wears a mask) or the factor $m$ (if both individuals in the interaction wear a mask). Caregiving interactions have a higher risk (by a factor $w_c$) than the baseline and weak interactions have a lower risk (by a factor $w_w$) than the baseline. Exposed individuals are not yet contagious; however, these individuals eventually transition to the asymptomatic state. From the asymptomatic state, an individual can either become symptomatically ill or be removed (R), which encompasses recovery, death, and any other situation in which an individual is no longer infectious. If an individual is symptomatic, they can either be removed or become hospitalized (H). From the hospitalized state, an individual eventually transitions to the removed state. The state-transition parameters that we have not yet mentioned are fixed rates of exponential processes.} \label{fig:compartments}
\end{center}
\end{figure}

We discretize time into units of $\Delta T = 1$ day. Our model, which one can view as an agent-based model, evolves as the individuals interact with other. Individuals who are in the S compartment can move into the E compartment, depending on their interactions on a given day. Each day, the probability that susceptible individual $i$ remains in the S compartment is
\begin{equation} 
	\sigma = \prod_{j \in B(i)} (1 - \beta w_w^{W_{ij}} w_c^{C_{ij}} m^{M_{ij}/2})\,, \label{eq:notget}
\end{equation}
where $B(i)$ is the set of all active (i.e., non-broken) contacts of individual $i$ {that are infectious}, $W_{ij}$ is $1$ if $i$ and $j$ are weak contacts and $0$ otherwise, $C_{ij}$ is $1$ if $i$ and $j$ have a caregiving relationship and $0$ otherwise, and $M_{ij}$ (which can be equal to $0$, $1$, or $2$) counts how many of individuals $i$ and $j$ wear a mask during an interaction between them. The term $\beta w_w^{W_{ij}} w_c^{C_{ij}} m^{M_{ij}/2}$ gives the probability that node $i$ becomes infected from an interaction with node $j$. Given $\sigma$, we compute the probability that $i$ transitions from the susceptible compartment (S) to the exposed compartment (E) in a given day:
\begin{equation} 
	\Pr(i \text{ transitions from } \text{S to E}) = 1 - \sigma \,. \label{eq:get}
\end{equation} 
We model the outcomes of transitions from E to A, transitions from A to I, transitions from A to R, transitions from I to H, transitions from I to R, and transitions from H to R as exponential processes with fixed rates of $\nu$, $\alpha$, $\eta$, $\mu$, $\rho$, and $\zeta$, respectively. In our simulations, transitions occur in intervals of size $\Delta T$ (which we set equal to $1$ day, as mentioned previously). When multiple transmissions between compartments are possible, such as from A to I and from A to R, we treat event transitions as competing exponential processes. (See Algorithms \ref{alg:overall}, \ref{alg:infect}, and \ref{alg:advance} of the Supporting Information.) We summarize the possible state transitions and their rates in Fig.~\ref{fig:compartments}. 

In our simulations, we uniformly-at-random initialize a fixed number $A_0$ individuals to be asymptomatic on day $0$. To account for limited testing availability in the early stages of the epidemic, we assume that only a fraction $\tau$ of individuals in the I compartment (i.e., they are symptomatically ill but not hospitalized) have a positive COVID-19 test. Having a positive COVID-19 test means that the individual has a {\it documented} case of COVID-19. We determine whether an individual will test positive if they are symptomatically ill at initialization when we assign them a true/false flag $P$ with probability $\tau$ for true. When $P$ is true, if that individual becomes symptomatic (I), we suppose (for simplicity) that they have a positive COVID-19 test immediately upon moving into the I compartment (i.e., before the next day begins). When $P$ is false, that individual only has a positive COVID-19 test if they are hospitalized. For simplicity, we again assume that their positive test takes place immediately upon moving into the H compartment. We also assume that we do not double-count individuals who have a positive COVID-19 test while in the I compartment if they later move into the H compartment. We assume that no asymptomatic infections are documented in the early spread of the disease. We compute daily tallies of cumulative documented cases at the end of each day. We need the assumption about not having positive COVID-19 tests of all infected individuals to be able to fit our parameters to the Ottawa data, which tabulates the number of documented cases (but not the cumulative number of total infections) over time.

In Table \ref{tab:params}, we present the parameters that we use in our model. We discuss and support the values of these parameters in Section \ref{sec:estimate} of our Supporting Information. Whenever possible, we seek to infer parameters directly from clinical data, instead of basing our parameters upon other models. (The exception is $\nu$, which is the transition rate from the E compartment to the A compartment.)

\begin{table}
	\begin{center}
		\begin{tabular}{c | p{2.in} | c | c | c}
			\bf Symbol & \bf Meaning & \bf Value & \bf Reference & Source \\
			\hline
			$f_\text{dis}$ & fraction of population who are disabled & $0.073$ & \cite{lauer20182017} & literature \\
			$f_\text{care}$ & fraction of population who are caregivers & $0.021$ & \cite{us_labor} & literature \\
			$f_\text{ess}$ & fraction of population who are essential workers & $0.1472$  &  \cite{essential_count,us_labor,us_pop} & literature \\
			$f_\text{gen}$ & fraction of population who are part of the general population & $0.7588$ & \cite{lauer20182017,us_labor,essential_count,us_pop} \\
			$m$ & risk-reduction factor from mask-wearing by both individuals in an interaction & $0.34$ & \cite{chu2020physical} & literature \\
			$b$ & probability that an ill individual breaks its weak contacts & $0.92$ & \cite{prbreak} & inferred \\
			$w_\text{w}$ & weak edge weight & $0.473$ & \cite{tian2020secondary,chu2020physical} & inferred \\
			$w_\text{s}$ & strong edge weight & 1 & N/A & by definition \\
			$w_\text{c}$ & caregiving edge weight & $2.27$ & \cite{madewell2020household,tian2020secondary} & inferred \\
			$\beta$ & baseline transmission probability & $0.0112$ & \cite{tian2020secondary} & inferred \\
			$\nu$ & transition rate from E to A & $1$ day$^{-1}$ & \cite{anderson2020will} & {borrowed} \\
			$\alpha$ & transmission rate from A to I & $0.0769$ day$^{-1}$ & \cite{buitrago2020occurrence,byrne2020inferred,ma2020epidemiological,hu2020clinical} & inferred \\
			$\eta$ & transmission rate from A to R & $0.0186$ day$^{-1}$ & \cite{buitrago2020occurrence,byrne2020inferred,ma2020epidemiological,hu2020clinical} & inferred\\
			$\mu$ & transmission rate from I to H & $0.0163$ day$^{-1}$ & \cite{bajema2020serious,jiehao2020case,byrne2020inferred,faes2020time} & inferred \\
			$\rho$ & transmission rate from I to R & $0.0652$ day$^{-1}$ & \cite{bajema2020serious,jiehao2020case,byrne2020inferred,faes2020time} & inferred \\
			$\zeta$ & transmission rate from H to R & $0.0781$ day$^{-1}$ & \cite{guan2020clinical} & inferred \\
			$\tau$ & probability of tested if ill but not hospitalized & $0.04$ & \cite{ottawa_case_counts} & fit \\
			$C^*$ & maximum number of contacts in power law & $60$ & \cite{ottawa_case_counts} & fit \\
			$\mathcal{D}_\text{strong}$ & distribution of strong contacts & $\mathcal{E}(0.283, 0.332, 0.155, 0.148, 0.0816)$ & \cite{census} & literature \\
			$\mathcal{D}_\text{pool}$ & distribution of pool sizes & $\mathcal{F}(10)$ & n/a & chosen \\
			$\mathcal{D}_{\text{ess,pre}}$ & essential worker weak-contact distribution when not distancing & $\mathcal{P}(0,C^*,21.37)$ & \cite{gallup} & inferred \\
			$\mathcal{D}_{\text{ess,post}}$ & essential worker weak-contact distribution during distancing & {$\mathcal{P}(0,C^*,21.37)$} & \cite{gallup} & inferred \\
			$\mathcal{D}_{\text{gen/dis,pre}}$ & general/disabled weak-contact distribution when not distancing & $\mathcal{P}(0,C^*,10.34)$ & \cite{gallup} & inferred \\
			$\mathcal{D}_{\text{gen/dis,post}}$ & general/disabled weak-contact distribution during distancing & $\mathcal{P}(0,C^*,7.08)$ & \cite{gallup} & inferred \\
			$\mathcal{D}_{\text{care,pre}}$ & caregiver weak-contact distribution when not distancing & $\mathcal{P}(0,C^*,5.14)$ & \cite{gallup} & inferred \\
			$\mathcal{D}_{\text{care,post}}$ & caregiver weak-contact distribution during distancing & $\mathcal{P}(0,C^*,4)$ & \cite{gallup} & inferred \\
			$A_0$ & number of asymptomatic individuals on day $0$ & $341$ & \cite{ottawa_case_counts} & fit \\
			$P_\text{Ottawa}$ & population of Ottawa & 994837 & \cite{ottawaPop} & literature \\			
			\hline			
		\end{tabular} \caption{The parameter values that we use in our study. In the ``Source'' column, {\it literature} means that we found a value directly {from data in} the literature; {\it inferred} means that we inferred a value based on published data in the literature; {\it by definition} signifies a value that we set in our model formulation; {\it chosen} indicates that a value is unknown, but we made a choice in our model; {\it borrowed} indicates that we adopted a value directly from a model in the literature; and {\it fit} indicates that we use Ottawa case data along with other (i.e., not fit) parameters in this table to estimate a value.}\label{tab:params}
	\end{center}
\end{table}

%%%%

\subsection{Summary of our Assumptions}
\label{sec:modelAssumptions}

We now briefly summarize the main assumptions of our model

\paragraph{Population:} Our model city's population is closed, so the city has no inflow or outflow.

\paragraph{Time units:} We discretize time in units of $\Delta T= 1$ day.

\paragraph{Composition:} The population of our city consists of the following types of individuals: $7.3$\% are disabled, $2.1$\% are caregivers, {$14.72$\%} are essential workers, and {$75.88$\%} are members of the general population. The roles of individuals do not change. 

\paragraph{Disease compartments:} Individuals can be susceptible, exposed, asymptomatic, symptomatically ill, hospitalized, or removed. All infected individuals must go through the exposed compartment before becoming infectious. 

\paragraph{Transitions between compartments:} We model an individual's daily infection rate through a probability of infection per interaction with a contagious individual, with interaction probabilities scaled up or down based on the types of interactions and the presence/absence of masks. All other transitions between compartments come from exponential processes that we compute one day at a time.

\paragraph{Strong contacts:} We assign the numbers of strong contacts of all individuals from the empirical probability distribution $\mathcal{D}_\text{strong}$. 

\paragraph{Weak contacts:} According to an individual's assigned role and the status of contact-limiting, we determine the numbers of their weak contacts from an approximate truncated power-law distribution (see Section \ref{sec:simulate}) using $\mathcal{D}_{\text{group,status}}$, where {\it group} is one of ``gen/dis'' (i.e., the general and disabled subpopulations), ``care'' (i.e., caregivers), or ``ess'' (i.e., essential workers) and {\it status} is one of ``pre'' (i.e., not during a lockdown) or ``post'' (i.e., during a lockdown). Recall that the weak-contact distribution has the same parameter values for disabled people and individuals in the general population.

\paragraph{Caregiving contacts:} All disabled people have a pool of weak-contact caregivers of a size that is dictated by $\mathcal{D}_\text{pool}$. For each disabled person, we choose that pool uniformly at random from the set of caregivers. Additionally, each disabled person has one strong caregiver that we choose uniformly at random from the set of caregivers and they see that individual each day, unless either the disabled person or that caregiver is hospitalized.

\paragraph{Breaking contacts:} Asymptomatic individuals do not break contacts (except in the form of contact-limiting). An ill (but not hospitalized) individual breaks their weak contacts with probability $b$. If an individual is hospitalized, they break both their weak contacts and their strong contacts until they move into the R compartment. An individual in the R compartment does not infect others with the disease; they may be deceased or simply no longer infectious. In our computations, those individuals regain their regular weak and strong contacts. 

\paragraph{Interactions:} Each day, an individual interacts with the same weak (except for caregiver--disabled interactions) and strong contacts unless the contact has been broken due to illness or when the contact distributions change. Each day, a disabled person interacts with their strong caregiver, unless illness prevents it. Each day, a disabled person interacts with a uniformly randomly selected member of their caregiver pool, unless illness prevents it. 
Even during contact-limiting stages, the weak contacts of essential workers do not break.

%%%%%

% RESULTS

%%%%%

\section{Results}
\label{sec:results}

\begin{figure}[!htbp]
        \centering
        \includegraphics[width=6.5in]{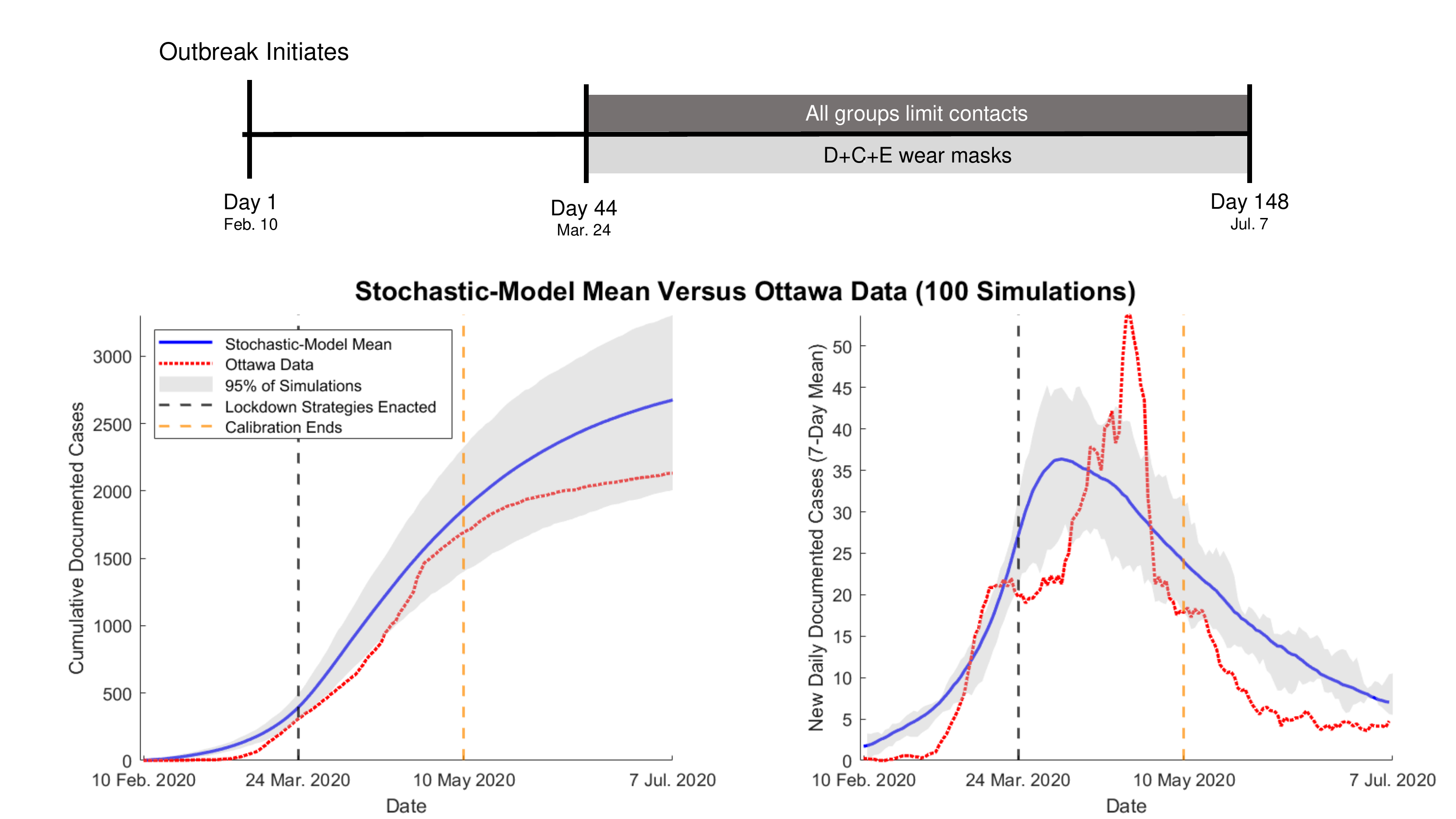}
        \caption{Comparison of a mean of 100 simulations of our stochastic model of COVID-19 spread with (left) cumulative documented case counts and (right) the 7-day mean of new daily documented cases. We calculate the 7-day mean is over a sliding window that includes the previous three days, the current day, and the next three days. We fit the parameters by minimizing the $\ell_2$-error of the model's predictions of daily case counts over the first 90 days. We show the mean of our model in blue and the Ottawa case data in red. The gray window indicates the middle 95\% of these 100 simulations. On day 44 (24 March, 2020), all subpopulations limit contacts and the (D+C+E) mask-wearing scenario begins. The graphs terminate on day 148, when Ottawa had its first reopening. }
        \label{fig:Ottawa}
    \end{figure}

We first compare the new daily documented cases and the cumulative number of documented cases in our model with empirical case data from Ottawa (see Fig.~\ref{fig:Ottawa}). We fit the parameters in our model up to May 10 (i.e., day 90) of the epidemic in Ottawa, and we assume that the city immediately enters a contact-limited phase on March 24 (i.e., day 44). We do the fitting (see Section \ref{sec:estimate} of the Supporting Information) by minimizing the $\ell_2$-error in the model's count of daily documented case counts. We show the 7-day mean of new daily documented cases; we calculate this mean over a sliding window that including the previous three days, the current day, and the next three days. At the endpoints, we truncate the window and take the mean over days that fill the window. We find reasonable agreement between the daily documented case counts in our model and the reported documented cases, but our match is not perfect. For example, the peak in daily documented case counts and the inflection point in cumulative documented cases occurs earlier in our model than it does in documented case records. This can arise from many possible factors, including delays in reporting cases (e.g., with differences on weekdays versus weekends), delays in the diagnosis of symptomatic individuals, changes of the model parameters (like testing availability) in time, or our use of only two degrees of freedom in our fits (with with most model parameters arising from sources that are not specific to Ottawa). The daily and cumulative documented case counts in our model deviate little from the data for the first $90$ days, but our model subsequently tends to overestimate the case count. We speculate that this may stem from overestimating the number of contacts of the Ottawans. Our contact estimates come from survey data \cite{gallup}, which do not focus specifically on Ottawa. We wish to avoid overfitting, so we accept the fit performance.

Additionally, there is large variance in epidemic trajectories; that is, the 95\% confidence window is large. We believe that one of the main factors behind this large variance is our use of an approximate truncated power-law distribution for weak contacts. If we replace these approximate truncated power-law distributions with deterministic distributions (i.e., distributions with $0$ variance) with the same mean values, we obtain much smaller variances, but the disease also does not spread. We have chosen to use truncated power-law distributions to allow large variations in numbers of contacts, but this results in a larger variance. See Section \ref{sec:further} of our Supporting Information for further discussion.

Our baseline transmission probability $\beta = 0.0112$ is smaller than those that were employed in some other studies \cite{pullano2021underdetection,Arenas2020}, which used $\beta \approx 0.06$. For the assumptions in our study, $\beta=0.0112$ is appropriate. With $\beta=0.06$, the disease is too infectious, and our simulations then result in a total documented case count that greatly exceeds the number of documented cases in Ottawa. This value of $\beta=0.06$ is also inconsistent with secondary attack-rate studies \cite{tian2020secondary} when they are combined with the durations that individuals spend in each compartment in our model. The fact that the disease can still spread so effectively with $\beta = 0.0112$ perhaps stems from our network structure, as some individuals can be superspreaders. 

\begin{figure}[!htbp]
        \centering
        \includegraphics[width=7.5in]{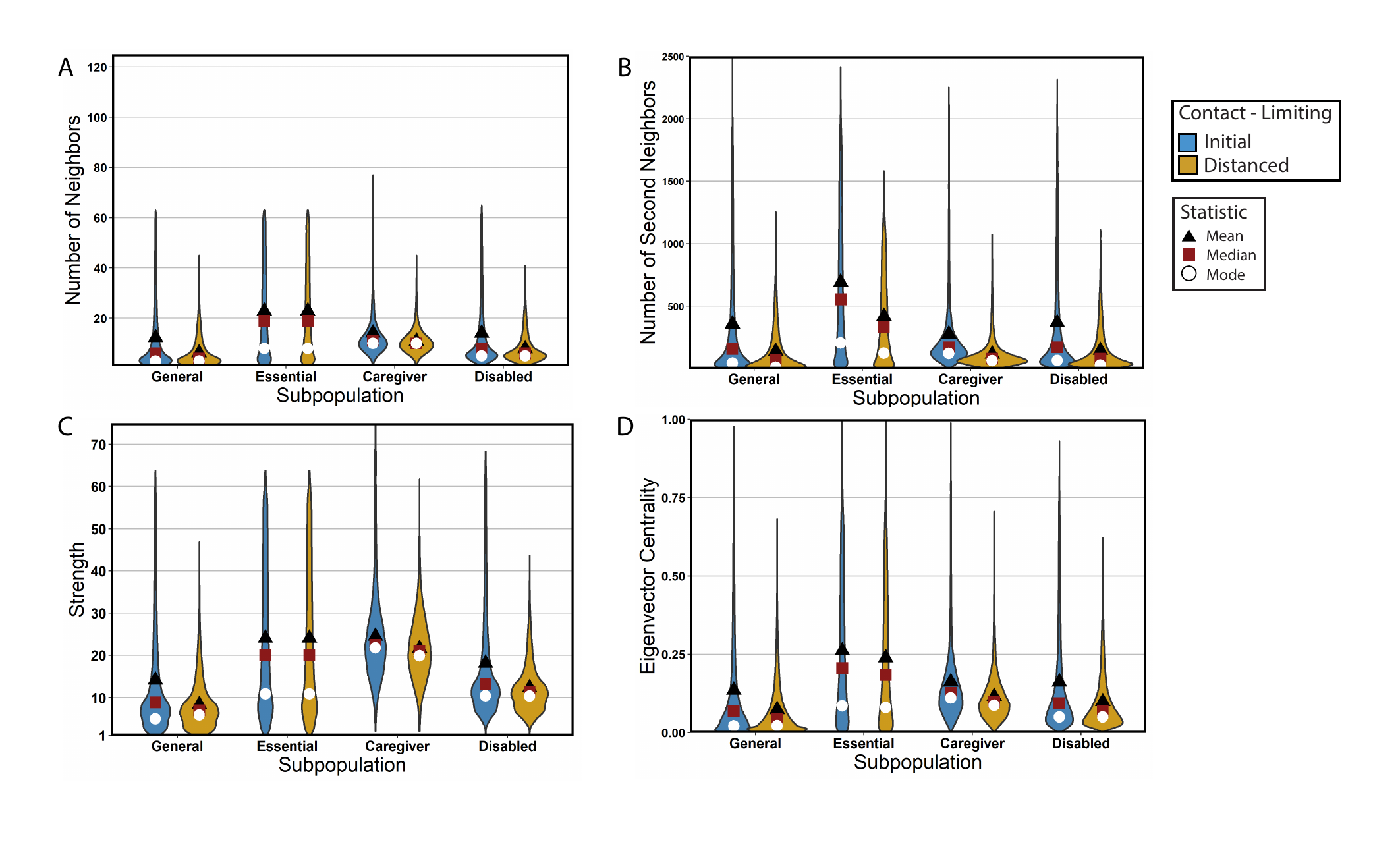} 
        \caption{Characterization of centrality measures of subpopulations in the networks on which we run our stochastic model of COVID-19 spread. The violin plots depict empirical probability densities. The initial situation, for which we show day 43 of one simulation, has no contact-limiting. The distanced situation, for which we show day 45 of the same simulation, has contact-limiting in all subpopulations. For each subpopulation, we calculate distributions of (A) the number of neighbors (i.e., direct contacts), (B) the number of second neighbors (i.e., contacts of contacts), (C) the strength of the contacts with neighbors, and (D) eigenvector centrality.
        }
        \label{fig:centrality}
    \end{figure}

\begin{figure}[!htbp]
        \centering
        \includegraphics[width=7.5in]{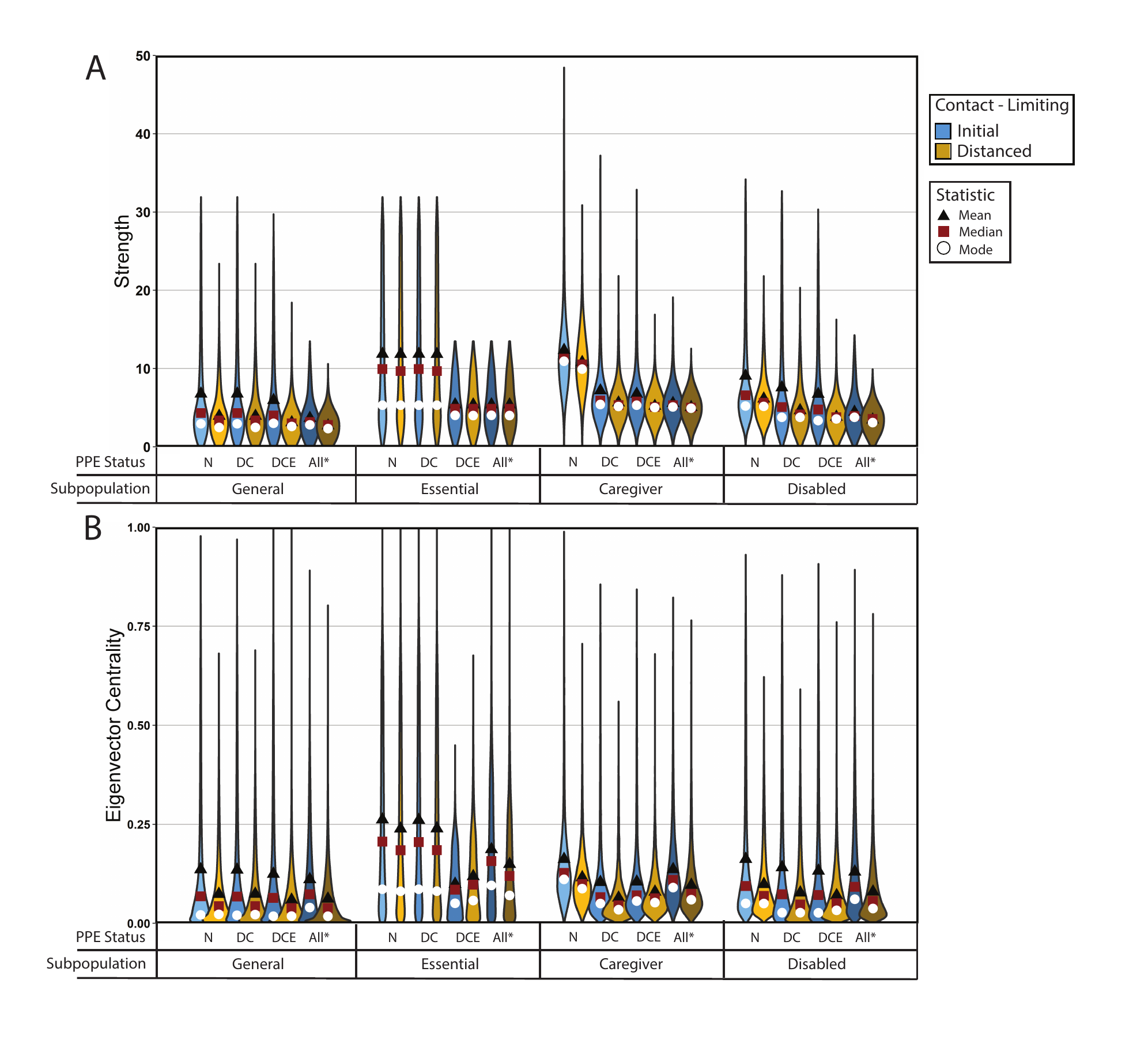} 
        \smallskip
        \caption{Characterization of the effects of mask-wearing on centrality measures of subpopulations in the networks on which we run our stochastic model of COVID-19 spread. The violin plots depict empirical probability densities. The initial situation, for which we show day 43 of one simulation, has no contact-limiting. The distanced situation, for which we show day 45 of the same simulation, has contact-limiting in all subpopulations. We modify edge strengths by supposing that masks have the effectiveness that we indicated in Table~\ref{tab:params}. To indicate the mask-wearing statuses of different scenarios, we use the notation that we defined in Section \ref{sec:modelSpecific}. For each subpopulation, we compute (A) the edge-weight distribution and the (B) eigenvector-centrality distribution.
        }
        \label{fig:PPEcentrality}
    \end{figure}

To help us understand the results of simulating our stochastic model of COVID-19 spread, we examine the structural characteristics of the networks on which we perform our simulations. Because different types of contacts have different levels of disease transmission, we base our measures on the statistics of weighted networks. Additionally, our network contact structure changes with time. For one network from our network model, we compare two days --- one before contact-limiting and one during it. For each of the two networks, we compute the number of first-degree contacts (i.e., direct contacts), second-degree contacts (i.e., contacts of direct contacts), contact strength (i.e., edge weight, which we interpret as the ``conductance'' of a disease across a contact), and eigenvector centrality (i.e., the leading eigenvector of the network's adjacency matrix, where larger values of eigenvector centrality are associated with ``high-traffic'' individuals, who are visited often by a random walker on the network \cite{newman2018networks,baek2020}). We are interested in eigenvector centrality because the largest eigenvector-centrality value in a network plays a role in determining that network's susceptibility to a widespread outbreak of a disease under certain conditions \cite{WangFaloutsos2003}. We consider the distribution of the eigenvector centralities for different subpopulations in our model city. We find that caregivers have the most first-degree and second-degree contacts (see Fig.~\ref{fig:centrality}A,B) and the largest mean strength (see Fig.~\ref{fig:centrality}C). Essential workers have the largest mean eigenvector centrality (because of the heavy-tailed distribution of their contacts), whereas caregivers have the largest modal eigenvector centrality (see Fig.~\ref{fig:centrality}D). We also test the effects of contact-limiting and mask-wearing (i.e., PPE status) strategies on the strengths and eigenvector centralities of various populations. Both NPIs reduce the strength of a node, and contact-limiting in particular reduces {the heavy tails of the distribution of strengths for caregivers, disabled people, and members of the general population (see Fig.~\ref{fig:PPEcentrality}).} In other words, contact-limiting reduces the probability that individuals have a large number of contacts.
  
We also test how much different contact-limiting and mask-wearing strategies affect the different subpopulations in our model. We consider different mitigation strategies, which we assume are deployed on day $44$, and we compare the of cumulative infections on day $148$ for these strategies. We consider the mask-wearing strategies {that we} outlined in Section \ref{sec:modelSpecific} and the following three contact-limiting strategies:
\begin{itemize}
    \item No contact-limiting: All people maintain their contacts for the entirety of the 148 days.
    \item Only disabled people limit their contacts. Disabled people reduce their number of weak contacts on day 44. All other populations maintain their contacts.
    \item Everyone except for essential workers limits contacts. All subpopulations other than essential workers reduce their number of weak contacts on day 44. 
\end{itemize}

We first consider the optimistic scenario in which all weak interactions involve mask-wearing. In this case, when everyone limits contacts on day 44, our simulations yield a mean of 13,242 cumulative infections through day 148. This is approximately 11.2\% lower than the 14,910 cumulative infections through day 148 when only caregivers, disabled people, and weak interactions with essential workers wear masks. We conclude that universal mask-wearing (specifically, in all situations except within households) is an effective NPI for reducing the number of COVID-19 cases. For all of our subsequent simulations, we assume that weak contacts wear masks only in interactions that involve essential workers , unless otherwise noted.

    \begin{figure}
        \centering
        \includegraphics[width=7.5in]{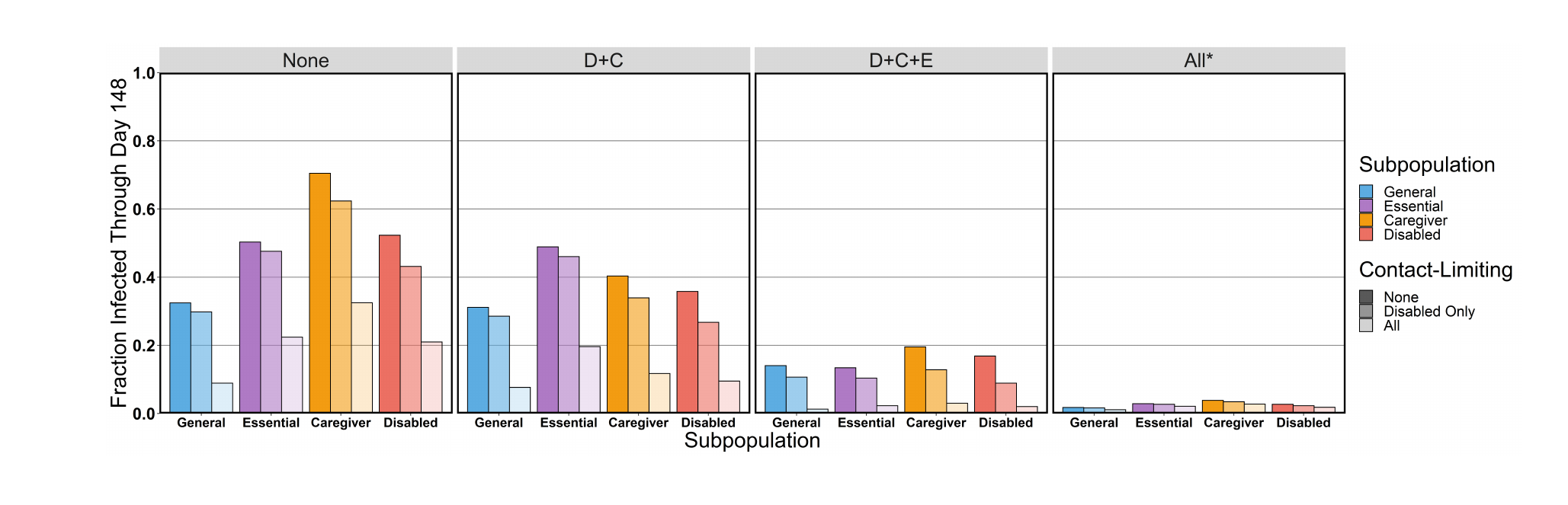}
        \caption{Mean cumulative infections in the general population (blue), essential workers (purple), caregivers (gold), and disabled people (red) for different contact-limiting and mask-wearing statuses. The mask-wearing statuses are the same as in Fig.~\ref{fig:PPEcentrality}.
        }
        \label{fig:MasksLimits}
    \end{figure}

We find that contact-limiting by only the disabled subpopulation has a relatively small effect on the number of their cumulative infections; it reduces the percent of them who become infected from 52.3\% to 43.1\%. Contact-limiting by only disabled people yields a similar result for caregivers, with a reduction in the percent of infected caregivers from 70.5\% to 62.4\%. Contact-limiting by all subpopulations has a larger effect; it reduces the percent of infected individuals in the disabled subpopulation to 21.0\% and that of caregivers to 32.5\%. Mask usage in both the disabled and the caregiver subpopulations protects both subpopulations even in the absence of any contact-limiting. The percent of disabled people who become infected decreases from 52.3\% to 35.8\%, and the percent of caregivers who become infected decreases from 70.5\% to 40.3\%. When essential workers, caregivers, and disabled people all wear masks, this protection is enhanced. The percent of disabled people who become infected decreases to 16.9\%, and the percent of caregivers who become infected decreases to 19.5\%. Finally, when all weak contacts wear masks, 2.7\% of disabled people and 3.8\% of the caregivers become infected. When all subpopulations limit contacts and all subpopulations wear masks (except within a household), 1.8\% of the disabled subpopulation and 2.7\% of the caregiver subpopulation become infected. We summarize our results of the mask-wearing interventions in Fig.~\ref{fig:MasksLimits}.

\begin{figure}
        \centering
        \includegraphics[width=4in]{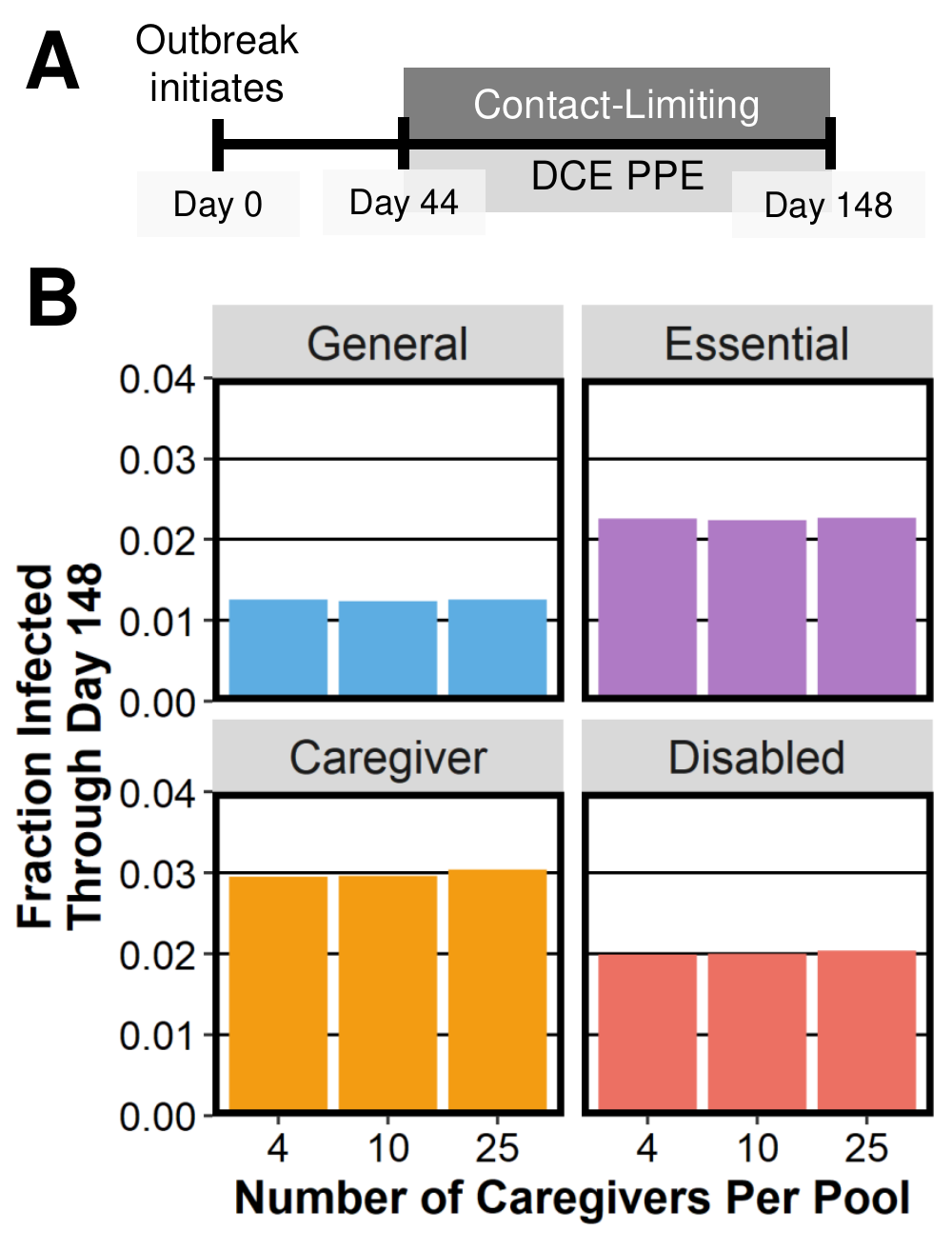}
        \caption{Effects of the number of caregivers (4, 10, or 25) 
        that are assigned to a given disabled person on the mean fraction of each subpopulation that becomes infected.
         The label ``DCE PPE" refers to the (D+C+E) mask-wearing scenario.
        }
        \label{fig:PoolSize}
    \end{figure}

Because COVID-19 guidelines recommend reducing the number of contacts between individuals, we test whether reducing the number of weak caregiver contacts per pool --- while maintaining daily caregiving interactions --- helps protect disabled people and/or caregivers. This NPI affects the total number of contacts of disabled people, but it does not reduce the total amount of time that they are exposed to these contacts. We test caregiver pool sizes of 4, 10, and 25, and we find that reducing caregiver pool size does not reduce infections among caregivers or disabled people (see Fig.~\ref{fig:PoolSize}). 
  
\begin{figure}
        \centering
        \includegraphics[width=4in]{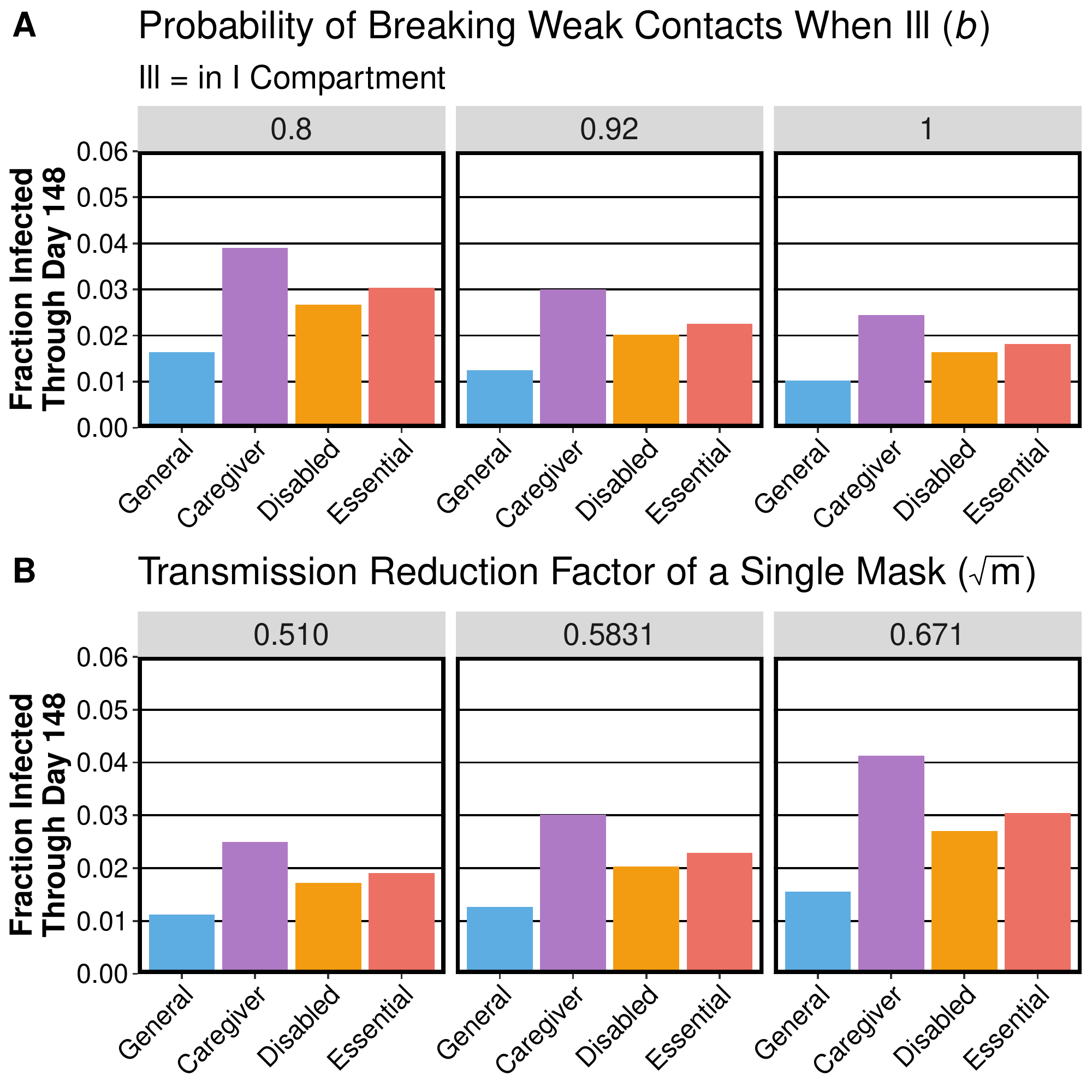}
        \caption{Effects of (A) the probability of breaking weak contacts when will and (B) mask effectiveness on the mean fraction that each subpopulation becomes infected.
        }
        \label{fig:VaryParameters}
    \end{figure}

In our investigation, we are particularly uncertain about the values of two parameters: the probability that individuals break weak contacts when they become ill and the effectiveness of masks. Therefore, we repeat our simulations with otherwise baseline conditions for different values of these parameters (see Table \ref{tab:params}). We choose the values of $m$ as the boundaries of the 95\% confidence window in mask effectiveness in \cite{chu2020physical}.We choose the values of $b$ as educated guess as to reasonable best-case and worst-case scenarios. As expected, reducing the probability of breaking weak contacts when ill (see Fig.~\ref{fig:VaryParameters}A) and reducing mask effectiveness (see Fig.~\ref{fig:VaryParameters}B) both increase the number of infections. Importantly, however, varying these parameters does not affect the overall pattern of infections; in particular, caregivers remain the most susceptible subpopulation.

\begin{figure}
        \centering
        \includegraphics[width=4in]{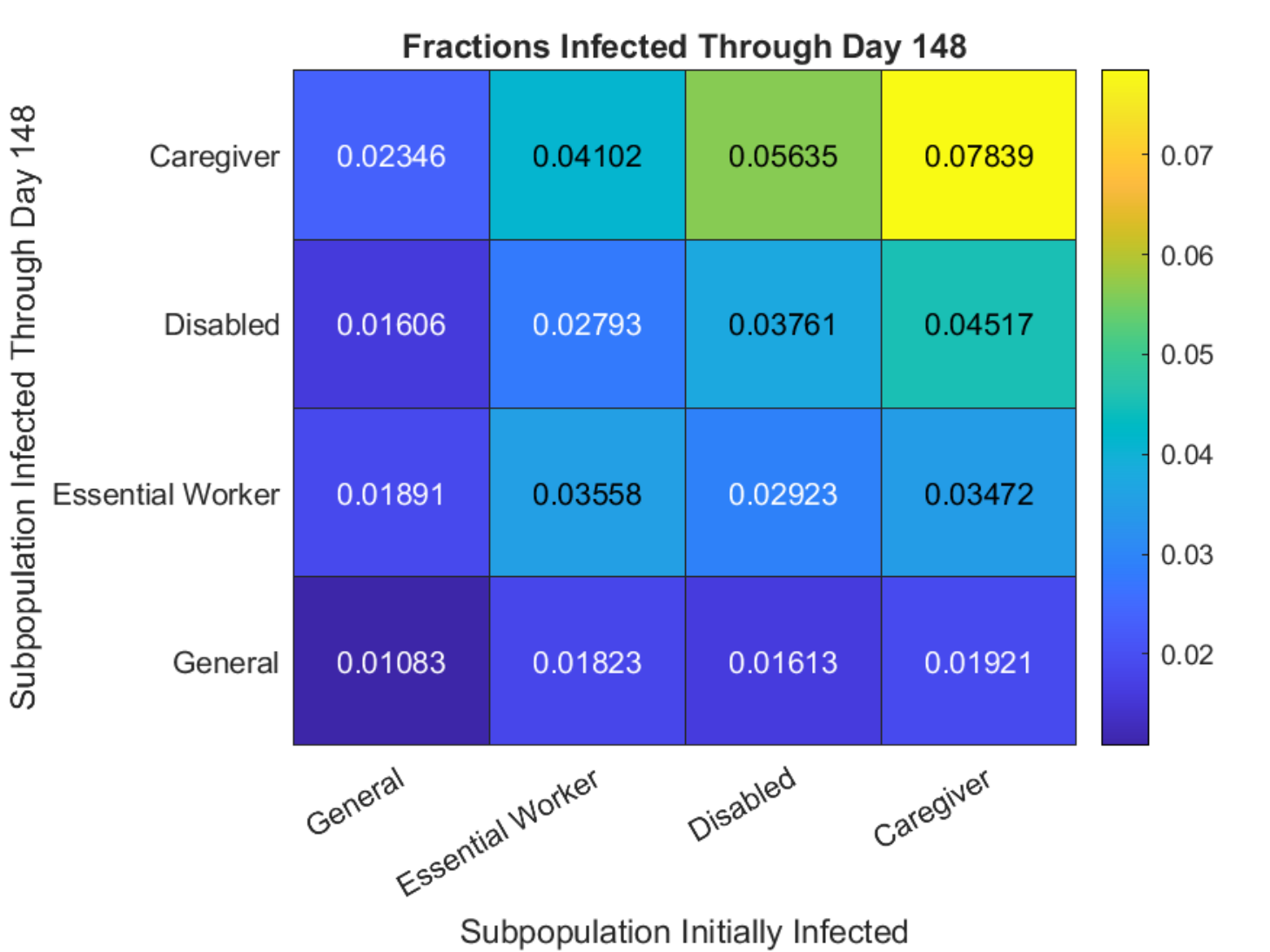}
       \smallskip
        \caption{Fraction of each subpopulation that is infected through day 148 when all of the initially infected individuals are in a single subpopulation. On day 44, all groups limit contacts and the (D+C+E) mask-wearing scenario begins.
   	}
        \label{fig:Seeding}
    \end{figure}

Having observed that caregivers are the most likely to be infected among all subpopulations across all tested parameter sets, we investigate whether caregivers are also the most prone to spreading COVID-19. To do this, we seed all initial infections entirely in a single subpopulation, rather than distributing the initially infected individuals uniformly at random across the entire population. We calculate the mean fraction of each subpopulation that was infected cumulatively through 148 days. We find that the caregivers are the most potent spreaders of COVID-19, with each subpopulation reaching its highest infection rate when only caregivers are infected initially (see Fig.~\ref{fig:Seeding}). 
Seeding all initial infections among only disabled people leads to the second-largest number of infections in the caregiver subpopulation.

As we explain in our Supporting Information (in Section \ref{sec:estimate}), because of the intimacy of caregiver--disabled interactions, the relative risk of such an interaction is likely higher than is the case for typical household interactions. In Section \ref{sec:further} of our Supporting Information, we also consider $w_c=1$ (i.e., the risk of a caregiver--disabled interaction is the same as a household interaction) $w_c=1.5$ (i.e., the risk of a caregiver--disabled interaction is only moderately higher than a household interaction). When $w_c=1$, 
essential workers are the most potent disease spreaders to all subpopulations except for the spread of the disease from caregivers to other caregivers. However, when $w_c=1.5$, caregivers are the most potent disease spreaders to the disabled community and to themselves, and essential workers are the most potent disease spreaders to the general population and to themselves. This suggests that our conclusions about the impact of the caregiver subpopulation on the disabled subpopulation are plausible even if the relative risk $w_c$ is only moderately larger than $1$.

\begin{figure}
        \centering
        \includegraphics[width=5.in]{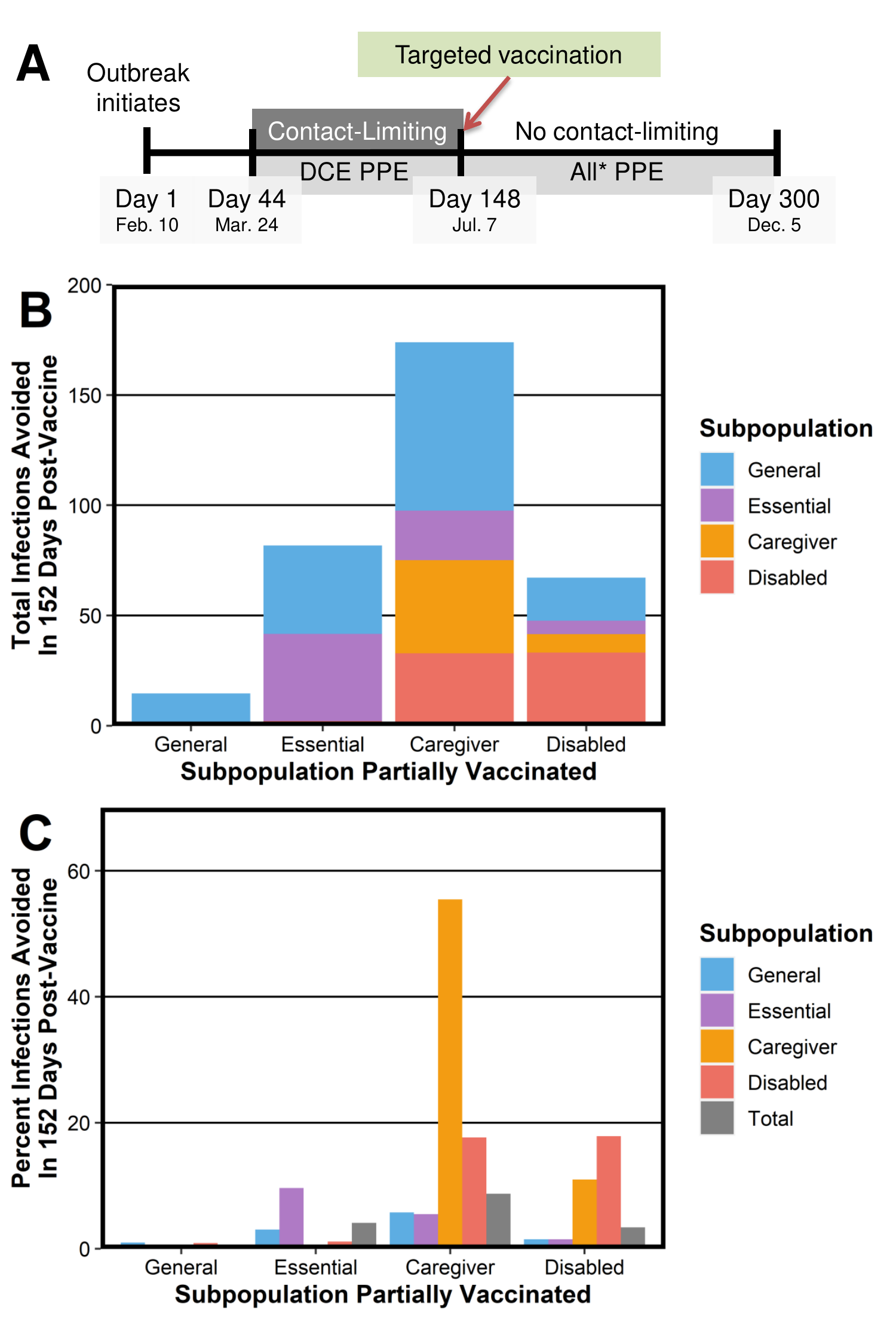}
        \caption{Number of infections that are prevented in each subpopulation when each subpopulation is vaccinated with the indicated number of vaccines.
        }
        \label{fig:StackVax}
    \end{figure}

Our finding that caregivers are the subpopulation that is most prone to spreading COVID-19 has potential implications for vaccine prioritization because vaccinating caregivers can indirectly protect other subpopulations. Because initial vaccine supplies are often extremely limited, we test the efficacy of providing vaccination to only a small fraction of the total population. To do this, we simulate the distribution of a very limited amount of vaccine --- equivalent to half (i.e., 10,151) of the mean remaining susceptible caregivers on day 148 (this is equal to approximately 1\% of the total city population) --- by moving a uniformly random subset of either susceptible caregivers, disabled people, essential workers, or the general population immediately to the removed compartment. When there were fewer susceptible people in a subpopulation than people to move, we move everyone in that subpopulation (and no other individuals) to the removed state. We also simulate a scenario with no vaccination. We simulate reopening at the same time as vaccination. In a reopening, all subpopulations return to their original weak-contact distributions, but all people wear masks during all non-household interactions. For a timeline, see Fig.~\ref{fig:StackVax}A. We then simulate our stochastic model of infections until day 300 and calculate the number of infections that are potentially preventable through the above vaccination strategies by comparing the results of these simulations to simulations that do not incorporate vaccination. This enables us to evaluate the benefits that vaccinating each subpopulation confers indirectly to other subpopulations.

Consistent with our previous findings, vaccinating caregivers {prevents} the largest number of infections. In our simulated scenario, targeting limited vaccinations to the caregiver subpopulation leads to a drop in total infections of 8.7\% in comparison to the scenario without vaccinations (see Fig.~\ref{fig:StackVax}B,C). It is second-most effective to vaccinate essential workers (this prevents 4.1\% of the total infections) and third-most effective to vaccinate the disabled subpopulation (which prevents 3.4\% of the total infections). Vaccinating the same number of individuals in the general population prevents only 0.7\% of the total infections. It is possible for the number of prevented infections to be smaller than the number of people who are vaccinated because not all vaccinated individuals would have become infected if they were not vaccinated. One plausible scenario in which this can occur is if the disease prevalence is relatively low in the subpopulation that is vaccinated, as is the case with the general population in our simulation. 

Vaccinating caregivers is an effective strategy to protect disabled people. When 10,151 caregivers are vaccinated, we reduce infections in disabled people by a mean of 17.7\%. Vaccinating the same number of disabled people spares a mean of 17.9\% of the disabled subpopulation (i.e., almost an equal number) from infection. These almost equal percentages may arise from the relative sizes of the caregiver and disabled subpopulations in our model. Vaccinating 10,151 individuals entails vaccinating exactly half of remaining susceptible caregivers, but 10,151 individuals constitutes only about 14\% of the disabled subpopulation. Therefore, when the number of vaccines is extremely limited, vaccinating caregivers may be comparably effective at protecting the disabled population as directly vaccinating disabled people.

Notably, vaccinating either the caregiver or the disabled subpopulation is much more effective at protecting the disabled subpopulation than vaccinating the essential-worker subpopulation, which only protects 1.1\% of the infections in the disabled subpopulation. Vaccinating caregivers also spares slightly more members of the general population than vaccinating essential workers; about 5.8\% of 
the general-population infections are prevented when 10,151 caregivers are vaccinated, and about 3.4\% of general-population infections are prevented when 10,151 essential workers are vaccinated. In our case study, the essential-worker subpopulation is the only subpopulation for whom the best strategy (among those that we considered) is to vaccinate the essential-worker subpopulation. With this strategy, they prevent 9.6\% of the infections, which is better than the
5.5\% that they prevent when the caregiver subpopulation is vaccinated (see Fig.~\ref{fig:StackVax}C). 

In our case study, we find that vaccinating the disabled subpopulation does not protect the caregiver subpopulation as effectively as vaccinating caregivers protects the disabled subpopulation. When 10,151 disabled people are vaccinated, a mean of about 11.0\% of the caregiver cases are prevented. When the same number of caregivers are vaccinated instead, about 55.5\% of the caregiver cases are prevented (see Fig.~\ref{fig:StackVax}). This fivefold difference may arise from the relative sizes of the caregiver and disabled subpopulations. Because a relatively small fraction of the disabled people with whom any given caregiver interacts will be vaccinated and caregivers are often in the pools of multiple disabled people, our case study suggests that caregivers' risks are not mitigated greatly when only a small fraction of the disabled subpopulation is vaccinated.

% DISCUSSION

\section{Discussion}
\label{sec:discussion}

We now summarize and discuss our key results.

%%%%

\subsection{Our Most Significant Findings}
{\bf The caregivers and disabled subpopulations are extremely vulnerable to COVID-19 infection}.
We simulated the spread of COVID-19 on networks to evaluate how vulnerable four interconnected populations --- caregivers, disabled people, essential workers, and the general population --- are to infection. Across multiple simulation conditions, we found that caregivers have the highest risk of infection and that disabled people have the second-highest risk of infection. This observation arises from multiple structural factors in our contact networks. First, there are many fewer caregivers than disabled people, so each caregiver typically has contact with multiple disabled people. This is reflected by caregivers having the largest number of direct neighbors and neighbors of neighbors. Second, caregiver--disabled connections are stronger than other connections, which --- along with the large number of direct connections of caregivers --- contributes to caregivers also having the edges with the largest mean weights. Third, some of our simulations involved a contact-limiting phase, in which individuals reduce their number of weak connections. However, caregiver--disabled contacts do not break during this phase. These structural factors render caregivers and disabled people particularly vulnerable to infection with COVID-19.  
We also found that caregivers are the most potent spreaders of COVID-19 once they are infected, and we suggest that this is due to the same factors {(specifically, being well-connected in a social newtork)} that make them most vulnerable to becoming infected. This agrees with the observations of Gozzi et al.~\cite{Gozzi2020}, who examined two different spread-limiting strategies in an activity-driven network model and found that the most active nodes that do not comply with a spread-limiting strategy are {the} major drivers of disease spread. Reassuringly, our findings are robust to changes in the parameters --- the effectiveness of masks and the probability of breaking contact when ill --- in which we had the greatest uncertainty. 

In our model, we assumed that the transition probability to become hospitalized is the same for all subpopulations, and we did not model death. However, disabled people are more likely than other individuals to have medical conditions that predispose them to severe cases of COVID-19 and accessibility barriers to receiving healthcare \cite{Kuper2021} and caregivers are more likely to belong to marginalized groups that are at increased risk due to systemic structural barriers in accessing medical care. Taking these factors into account may reveal an even more disproportionate disease burden on caregivers and disabled people. Ortega-Anderez et al.~\cite{OrtegaAnderez2020} observed that small decreases in the exposure of medically vulnerable subpopulations significantly decrease overall mortality, underscoring how critical it is to identify interventions that effectively protect caregivers and disabled people. 

%%%%

\medskip

\textbf{Effective interventions}.
It is essential that the necessary medical services that at-home caregivers provide to disabled people continue to be available during a pandemic. These services are essential for survival; going without caregiving services endangers a disabled person's health. Therefore, we tested the effectiveness of various NPIs at preventing the spread of COVID-19 among these subpopulations. We found that mask-wearing during contact between caregivers and disabled people is a very effective strategy for reducing infections in both subpopulations. This finding agrees with recent agent-based \cite{Bahl2020, LiMinhas2020} and bond-percolation \cite{TianPoor2020} models of mask-wearing interventions. We recommend that home-healthcare agencies provide their employees with masks and (whenever possible) mandate their use on the job. 

Additionally, we found that contact-limiting by disabled people alone only slightly reduces their risk of contracting COVID-19 if it is not accompanied by contact-limiting in the rest of the population. When all groups limit contacts, cases among disabled people and caregivers fall by almost 50\%. {This result underlines the fact that changes in behavior in the general population can drive changes in disease spread in the disabled subpopulation. Disabled people alone are not numerous enough to change large-scale epidemic dynamics with their behaviors, and they are vulnerable to increases in disease spread that can occur when the general population changes its behavior. In the context both of the current COVID-19 pandemic and possible future pandemics, we emphasize the critical influence of behavior by the general popular on disabled communities. Mitigation efforts by the general population, such as contact-limiting (as in the present study), can protect disabled people much more than interventions in only the disabled subpopulation.

%%%%

\medskip

\textbf{Vaccinating caregivers shields other subpopulations, including disabled people.}
A major application of modeling of the spread of a disease on a network is evaluating strategies for targeted vaccination \cite{WangReview2016,PastorReview2015,kiss2017}. Prior research suggests that, under certain conditions, the largest eigenvector centrality of a network helps determine a network's threshold (e.g., in the form of a basic reproduction number) for a widespread outbreak of a disease \cite{WangFaloutsos2003}. This suggests that vaccinating nodes with large eigenvector centralities may be an effective control strategy. Several COVID-19 vaccines have received rigorous safety testing for approval across the globe \cite{USAVax, EuroVax, CanadaVax, UKVax}, and we sought to determine the most effective vaccination strategy in the context of our model. As a first step, we calculated eigenvector centralities of the nodes in the network's four subpopulations. We calculated that essential workers have the largest mean eigenvector centralities in a single simulated population and that caregivers have the largest modal eigenvector centralities in the same simulated population. The difference between the mean and modal values is a direct consequence of the contact distributions of these two subpopulations. For example, essential workers are sometimes in very large workplaces and sometimes in very small workplaces, whereas caregivers almost always work with multiple disabled people. 

Investigating network structure alone in our model did not resolve which subpopulation is the most efficient one to vaccinate. Therefore, we analyzed how the dynamics of disease spread were affected by selectively vaccinating a subset of each of these subpopulations. We considered a hypothetical vaccine that is completely effective and permanently prevents any individual who receives it from contracting or spreading the virus SARS-CoV-2. Although this is unrealistic --- vaccinated people can still contract SARS-CoV-2 and even spread it to others \cite{BrownMMWR2021} --- vaccinated people are much less likely than unvaccinated people to be diagnosed with the disease COVID-19 \cite{KaiserFamily2021}. They also experience a faster drop in viral load when they are infected, so transmission periods may be shorter in vaccinated people \cite{Chia2021}. 

Vaccine effectiveness against household transmission of so-called ``breakthrough cases'' of COVID-19 infection in vaccinated individuals was estimated at 71\% in one study \cite{deGier2021}. However, this study was conducted when the Alpha variant (Pango lineage designation B.1.1.7) of SARS-CoV-2 was predominant, and it is unknown whether this finding holds for the Delta variant (Pango lineage designation B.1.617.2). As new variants emerge frequently and vaccine adherence, availability, and manufacturers vary worldwide, we chose to examine a simplistic scenario instead of attempting to model any specific real-world situation. 

We measured the effectiveness of vaccine strategies by comparing the number of infections in scenarios with and without vaccination (without minus with). This number includes both infections that are prevented directly (specifically, when an individual who would have become infected had already received the vaccine) and ones that are prevented indirectly (specifically, chains of transmission that did not occur because individuals who would have spread the virus were instead vaccinated against it). Our simulations suggest that vaccinating caregivers (1) prevents the largest total number of infections and (2) prevents the most infections in three of the four subpopulations. (The exception is the subpopulation of essential workers.) In our simulations, vaccinating a specified number of caregivers protected an almost equal number of disabled people from infection (because of indirect prevention) as vaccinating the same number of disabled people.

It is necessary to be cautious when interpreting our findings about the relative efficiency of vaccinating various subpopulations. To obtain our results, we assumed that vaccines prevent the spread of COVID-19 from a vaccinated individual to other individuals. In the extreme hypothetical scenario where vaccines prevent serious illness but have no effect on viral transmission from vaccinated individuals, it is likely better to employ them in populations (e.g., disabled people) that are more likely to experience hospitalization and death. Moreover, even if vaccinating caregivers does turn out to be the most efficient way to reduce total case numbers of COVID-19, it may still be more ethical to prioritize vaccinating individual disabled people, particularly those who are elderly or have conditions that predispose them to severe disease \cite{OrtegaAnderez2020}. In the real world, vaccination campaigns must balance many factors --- including medical risk, public health, and equity --- when assigning priority \cite{Toner2020}. Additionally, we reiterate that the precise conclusions about vaccination strategies from our model may not hold in real-world scenarios. For example, it is important to consider a variety of local factors, including the amount of vaccine that is available, the relative sizes of the caregiver and disabled populations, and the distributions of age and pre-existing conditions in these populations. 
 
When a small number of caregivers serve a large number of disabled people who are not at particularly high medical risk, vaccinating caregivers has several benefits: (1) it protects caregivers, who often are in demographic groups with an elevated risk of COVID-19 complications, for their own sake; (2) it indirectly shields the disabled people for whom they care; and (3) it prevents the disruption of essential caregiving services to disabled people when caregivers are infected and must quarantine. Furthermore, for disabled people who cannot gain the benefit of vaccination --- whether due to access issues with transportation or at vaccination centers, immunosuppression, or other health challenges --- our findings suggest that vaccinating caregivers can be an extremely useful preventative strategy.

Our model strongly suggests that caregivers of disabled people are at increased occupational risk of both contracting and spreading COVID-19 and that protecting caregivers also provides substantial, quantifiable benefits to the vulnerable population that they serve. Therefore, we suggest that caregivers should be among the groups that are given the opportunity to receive a vaccine as a high priority. 

Especially when vaccines are not readily available, we emphasize the importance of continuing effective NPIs, such as mask-wearing and contact-limiting, in all subpopulations (including the general population). Additionally, vaccination campaigns should make it a priority to protect disabled people, and they should consider early vaccination of caregivers and disabled people as one potential strategy among continued society-wide NPIs to accomplish this goal. 

%%%%

\subsection{Limitations of our Study}

In interpreting our results, we made many assumptions to construct a tractable model to study. Accordingly, our results occur in the context of a variety of hypotheses about the epidemiology of COVID-19 in the disabled community and optimal strategies to mitigate the spread of the disease. Although we consider our hypotheses to be reasonable ones, we obviously did not perfectly describe the complexity of COVID-19, how it spreads, and how human behavior affects its spread. (See \cite{Bedson2021} for a recent review agenda for integrating social and behavioral factors with disease models.)

In reflecting on our assumptions and our modeling (of both network structure and the spread of COVID-19), there are a variety of natural steps to take to enhance our work. Although they are beyond the scope of the present paper, we elaborate on some of them. We encourage careful examination of the following ideas:
\begin{itemize}
\item Incorporating skilled nursing facilities and hospitals: We assumed that caregivers provide at-home care to disabled people. There are many disabled people who live in skilled care facilities, which have a different network of care-giving and care-receiving than the ones that we examined. One can perhaps argue that we modeled these effects indirectly through the heavy-tailed distributions of weak contacts, but that does not incorporate the intricacies of these healthcare settings. 

\item Lack of entry into and exit from a city: We did not consider the possibility that people enter a city, which can introduce more infections. This type of effect was studied in \cite{lai2020}.

\item Uncertainty in the numbers of disabled people and caregivers in a population: There is a lot of uncertainty in the proportions of disabled people and caregivers in a population. Unfortunately, there is not much reliable information about how many disabled people receive assistance for their activities of daily living and how many people in society serve as caregivers (possibly in an unpaid or undocumented capacity). It is very important to obtain more data about this and to incorporate it into modeling efforts.

\item More precise distributions of contacts: It was very difficult for us to estimate the contact distribution of people before and during a lockdown, and it was even more difficult to estimate the level of contact-limiting. It is worthwhile to study the effects of different types of distributions of weak contacts. We briefly explore this issue in Section \ref{sec:further} of the Supporting Information.

\item Incorporating daily randomness of interactions: During each phase of our model COVID-19 pandemic, we fixed the set of potential daily contacts (they are only potential contacts because illness can temporarily sever ties) of the population's individuals, except for interactions between disabled people and caregivers, for whom we assigned a random caregiver from a pool to each disabled person.

\item Modeling contact changes during a city's reopening: One limitation of our network model is that when we assigned additional contacts to individuals after our model city reopens, we did so in a random way (for simplicity), rather than having individuals resume the contacts that they had before a lockdown. This choice mixes the contacts in the network, and studying the consequences of this choice seems important.

\item Heterogeneity in mask effectiveness: We assumed that all masks give the same transmission-reduction benefits. However, this is not realistic. There are a large variety of mask types and some people do not wear masks correctly, so it seems worthwhile to examine how heterogeneity in mask effectiveness affects disease dynamics.

\item Modeling mask-compliance probabilistically: We assumed that masks are either worn or not worn by an entire subpopulation for given types of interactions. In reality, only some fraction of a subpopulation will wear masks.

\item Studying the importance of caregivers to disease spread: We speculated that the larger modal eigenvector centralities of caregivers leads to COVID-19 spreading more effectively from caregivers than from other subpopulations. It seems useful to study the importance of caregivers to disease spread from a theoretical perspective. For example, given the properties of the contact distributions of different members of a society, it is desirable to investigate the probability with which a large-scale epidemic occurs and how quickly it spreads in its early stages if only caregivers are infected when it starts.

\item Modeling vaccination outcomes: Our model simplistically assumes that vaccination fully prevents COVID-19 infection and transmission. In reality, vaccination provides robust but incomplete protection from COVID-19. Vaccinated individuals can experience asymptomatic or symptomatic disease and can transmit the virus to others, although at lower rates than unvaccinated individuals \cite{LACounty2021}. Our model does not take into account infection of or transmission by vaccinated individuals.

\item Effects of new variants of SARS-CoV-2: The SARS-CoV-2 virus has mutated over time, and some of our parameter estimates may depend on specific strains of the virus and differ across both time and geographic regions.

\item Uncertainties in timing: We used the simplistic assumption that all positive tests of COVID-19 of individuals in the I and H compartments occur at the beginning of an individual's first day in the relevant compartment. We also assumed that the availability of COVID-19 tests was the same throughout the first 148 days of the COVID-19 pandemic. Neither of these assumptions is realistic, and it seems worthwhile to consider more realistic testing scenarios.

\end{itemize}

%%%%%

\subsection{Conclusions}

We constructed a stochastic compartmental model of the spread of COVID-19 on networks that model a city of approximately $1$ million residents and used it to study the spread of the disease in disabled and caregiver communities. Our model suggests that (1) caregivers and disabled people may be the most vulnerable subpopulations to exposure in a society (at least among the four subpopulations that we considered); (2) mask-wearing appears to be extremely effective at preventing infections in caregivers and disabled people; (3) contact-limiting by an entire population appears to be far better at protecting disabled people than contact-limiting only by disabled people; and (4) caregivers may be the most potent spreaders of COVID-19 and vaccinating caregivers can be extremely helpful to protect disabled people.

%%%%%%

\section*{Acknowledgements}

We gratefully acknowledge Deanna Needell and Sherilyn Tamagawa for making the introductions that allowed our team to form, and we thank Stephen Campbell (Data and Policy Analyst at PHI)
for directing us to helpful resources and helping refine our questions. 
MAP acknowledges support from the National Science Foundation (grant number DMS-2027438) through the RAPID program.

\section*{Competing interests}

HS provides care, and JZ and ST receive care. ST reports on the COVID-19 pandemic as a journalist.

%%%%
\bibstyle{vancouver}

%%%%

%%%%

\section*{Supporting Information}

\appendix

% PARAMETERS

%%%%%

%%%

\section{Estimates of Parameter Values}\label{sec:estimate}

We present the assumptions and derivations that we use to estimate the parameters in our model (see Section \ref{sec:model}). These include parameters that we can obtain directly (possibly with some inference) from the literature and ones that we fit from case data in Ottawa. In this section, we use $\log$ to denote the natural logarithm.

%%%%

\subsection{Parameters that we Infer from the Literature} \label{infer}

\subsubsection{Properties of Exponential Distributions}

Because we assume that transition times between disease states come from exponential distributions, we state a few useful properties of exponential random variables. 

For a random variable $X$ that one samples from an exponential distribution with rate $\lambda$ (i.e., $X \sim \text{Exp}(\lambda)$), the probability density function is $f(x) = \lambda \e^{-\lambda x}$, the mean is $1/\lambda$, and the median is $\log 2/\lambda$. 

Suppose that we have a random variable $Y = \min \{ Y_1, Y_2 \}$, where $Y_1$ and $Y_2$ are random variables that we sample from exponential distributions of rates $\lambda_1$ and $\lambda_2$, respectively. It then follows that $Y$ is an exponential random variable with rate $\lambda_1 + \lambda_2$ and the probability that $Y_1 < Y_2$ is $\lambda_1/(\lambda_1 + \lambda_2).$

%%%%

\subsubsection{Parameters}

\paragraph{Transition Rate from Exposed to Asymptomatic ($\nu$).} The rate of moving from the exposed compartment to a contagious state (and hence to the asymptomatic compartment in our model) has been estimated at $\nu = 1$ day$^{-1}$ \cite{anderson2020will}.

\paragraph{Recovery Rate from Hospitalization ($\zeta$).} The mean duration of hospitalization has been estimated to be ${1}/{\zeta} = 12.8$ days \cite{guan2020clinical}, so $\zeta \approx 0.0781$ day$^{-1}$.

\paragraph{Transition Rates from Asymptomatic to Ill ($\alpha$) and Recovered ($\eta$).} It has been estimated that $19.45$\% of cases are entirely asymptomatic \cite{buitrago2020occurrence}, so $\frac{\eta}{\eta+\alpha} = 0.1945$. Byrne et al.~\cite{byrne2020inferred} summarized many relevant studies that give data about different transition rates. From these studies, we note that the mean duration in the asymptomatic state has been estimated to be about $\frac{1}{\alpha+\eta} = 7.25$ days \cite{ma2020epidemiological} and the median duration 
has been estimated to be about $\frac{\log 2}{\alpha+\eta}=9.5$ days \cite{hu2020clinical}. We take the mean of these two values to estimate $\frac{1}{\eta+\alpha} \approx 10.478$ days, which we combine with $\frac{\eta}{\eta+\alpha} = 0.1945$ to obtain $\alpha \approx 0.07688$ day$^{-1}$ and $\eta \approx 0.01856$ day$^{-1}.$

\paragraph{Transition Rates from Ill to Hospitalized ($\mu$) and Recovered ($\rho$).} It has been estimated that approximately $\frac{\mu}{\mu+\rho} = 20$\% of the symptomatic cases of COVID-19 result in hospitalization \cite{bajema2020serious}. Among children with mild cases of COVID-19, the median duration from the onset of symptoms onset no longer being infectious is about $\frac{\log 2}{\mu + \rho} = 12$ days \cite{jiehao2020case}. (This study was also referenced in Byrne et al.~\cite{byrne2020inferred}.) In Belgium, the median duration from the onset of symptoms to hospitalization was estimated to be $\frac{\log 2}{\mu + \rho} = 5$ days \cite{faes2020time}. We take the mean of the values from these two studies and estimate $\frac{1}{\mu+\rho} \approx 12.2629$. With $\frac{\mu}{\mu+\rho} = 0.2$, we obtain $\mu \approx 0.01631$ day$^{-1}$ and $\rho \approx 0.06524$ day$^{-1}.$

\paragraph{Mask Risk-Reduction Factor ($m$).} Based on three different viruses (SARS CoV-2, SARS-CoV, and MERS-CoV), an unadjusted relative risk when wearing a face mask versus not wearing one and contracting an infection has been reported to be $0.34$ (with a 95\% confidence window of $0.26$ to $0.45$) \cite{chu2020physical}. These results include both healthcare settings and non-healthcare settings. Because the three viruses are from the same family, it was argued in \cite{chu2020physical} that their relative risks should be comparable. 
For the data that they reported, it is not clear if only one or both individuals wore masks in their interactions. We use $m = 0.34$ to represent the risk reduction when both individuals in an interaction wear masks, and we use $\sqrt{m} \approx 0.5831$ if only one individual in an interaction wears a mask. That is, if only one individual in an interaction wears a mask, we quantify the transmission risk as the geometric mean of the best-case transmission reduction if both individuals wear a mask and the worst-case transmission reduction if neither individual wears a mask. By definition, given values $q_1, q_2, \ldots, q_n$, their geometric mean is $(q_1 \times q_2 \times \cdots \times q_n)^{1/n}$. Although our choice seems arbitrary, according to \cite{brainard2020community}, there is a small reduction in the chance of becoming infected in people who wear masks within a household, so it seems plausible that one individual wearing a mask in an interaction between two people confers some reduction in transmission.
 
\paragraph{Probability of Breaking Weak Contacts if Symptomatic ($b$).} It was very difficult to estimate this parameter. Ultimately, we use the fact that $92$\% of people in a survey reported practicing physical distancing \cite{prbreak} as a proxy for the portion of a population who would break their weak contacts if they became symptomatic. That is, $b = 0.92$.

\paragraph{Baseline Transmission Probability $\beta$ and Caregiving ($w_\text{c}$) and Weak ($w_\text{w}$) Edge Weights.} We estimate $\beta$ and these edge weights based on reported secondary attack rates in various scenarios. The secondary attack rate describes the fraction of a contagious individual's contacts who become infected as a result of interacting with that individual. The secondary attack rate among weak contacts \cite{tian2020secondary} appears to range from about 1\% to about 6\%, so we estimate it to be $3.5$\%. Additionally, the secondary attack rate within a household has been estimated to be approximately $20\%$ \cite{tian2020secondary} and is much higher (about $37.8$\%) between spouses \cite{madewell2020household}.

Caregiving work is extremely intimate and requires extended, close physical contact and potential exposure to bodily fluids. Such a level of intimacy is not typical between housemates, so we use the secondary attack rate between spouses as a proxy for the level of risk in an interaction between a caregiver and a disabled person.

We conduct a set of simulations to estimate the secondary attack rate for each type of contact. The secondary attack rate is the fraction of a contagious individual's contacts that they infect on average. In each trial, we assign a contagious duration $D_c$ (asymptomatic time plus possibly symptomatic time, depending on contact type and whether contacts are broken if an individual becomes ill) and compute the probability that that contagious individual infects somebody. For weak contacts\footnote{We obtain the value $\sqrt{m}$ by estimating the risk mitigation of masks as the geometric mean of the value ($1$) when no individual in an interaction wears a mask and the value ($m$) when both individuals in an interaction wear a mask. We use the geometric mean because of the uncertainty in whether or not people wear masks.}, we use a daily transmission probability of $\sqrt{m} w_w \beta$; for strong contacts, we use a probability of $\beta$; for caregiving contacts, we use a probability of $w_c \beta$.

Therefore, in a single trial, the probability of infection via a strong contact is $1-(1-\beta)^{D_c}$. We then determine the values of $w_w$, $\beta$, and $w_c$ so that when averaged over many trials, the mean probability of passing on COVID-19 matches the above secondary attack rates. This yields $\beta \approx 0.0112$, $w_w \approx 0.473$, and $w_c \approx 2.268$.

\paragraph{Subpopulation Proportions of the Total Population.} By combining the fraction of the population that has a cognitive disability with the fraction that has a physical disability that causes difficulty in dressing, bathing, or getting around inside the home, we estimate that the fraction of our population who are disabled and receive assistance from professional caregivers is $f_\text{dis} \approx 0.073$\cite{lauer20182017}. Unfortunately, there is a paucity of readily available data, so this is a rough estimate. From the United States Bureau of Labor Statistics, a fraction $f_\text{care} \approx 0.021$ of the U.S. population is employed as a home health/professional care aid \cite{us_labor}. We use this number as an estimate of the proportion of the population that provides care. This is likely an underestimate because many people provide care in unpaid settings. From an estimated 55,217,845 essential workers in the United States \cite{essential_count}, whose population in July 2019 was estimated to be 328,239,523 \cite{us_pop}, the fraction of essential workers  is approximately $0.1682$. After subtracting the people who are caregivers, we obtain that a fraction $f_\text{ess} \approx  0.1472$ of the population are essential workers. That leaves the fraction $f_\text{gen} \approx 0.7588$ for the remaining population (i.e., the general population).

\paragraph{Mean Numbers of Contacts.} We need distributions of the numbers of family contacts, weak contacts (through work, shopping, seeing friends, and so on), and caregiving contacts. We begin by focusing on the mean values and later consider the distributions themselves. From the 2016 Canadian census \cite{census}, households have a mean value of $2.4$ members, which implies that individuals have a mean of $\bar F = 1.4$ strong contacts.

From Gallup data in April 2020 \cite{gallup}, during pandemic lockdowns, the people who were surveyed had a mean of $5.1$ contacts per day at work and a mean of $4$ contacts per day outside of work and home. Additionally, 27\% of working adults completely isolated themselves except to members of their own household. In Europe in 2008, the overall population had a mean of $13.4$ daily contacts without a lockdown in place \cite{mossong2008social}. In April 2020, essential workers saw a mean of $22$ contacts per day (a much larger number than people who are not essential workers) during the lockdown \cite{rothwell_2020}. By combining these disparate pieces of data, we are able to make some relevant estimates.

Let $O_\text{gd}$ denote the mean number of occupational contacts of the general and disabled subpopulations each day without a lockdown, $O_\text{c}$ denote the mean number of disabled people that a caregiver sees in a day, $O_\text{gd}^*$ denote the mean number of occupational contacts of the general and disabled subpopulations each day during a lockdown, $w$ denote the mean number of weak contacts (outside of work) of
the overall population each day without a lockdown, $w^*$ denote the mean number of weak contacts (outside of work) of the overall population each day with a lockdown, and $O_\text{e}$ denote the mean number of occupational contacts of essential workers each day (both with and without a lockdown). Our parenthetical comment about $O_\text{e}$ indicates that we are assuming that the number of work contacts is the same for essential workers regardless of whether or not there is a lockdown. We also assume that $w$ does not depend on an individual's subpopulation (disabled person, caregiver, essential worker, or member of the general population). Likewise, we assume that $w^*$ does not depend on an individual's subpopulation.

From the data that we cited two paragraphs ago, we estimate that $w^* = 4$ and each disabled person sees $2$ caregivers per day. Additionally, $O_\text{c} = \frac{2 f_\text{dis}}{f_\text{care}} \approx 6.95$ and
\begin{align*}
	22 &\approx \frac{f_\text{ess} (O_\text{e} + w^* + \bar F) + f_\text{care}(O_\text{c} + w^* + \bar F)}{f_\text{ess} + f_\text{care}} \\
5.1 &\approx (f_\text{care} + f_\text{dis}) O_\text{gd}^* + f_\text{ess} O_\text{e} + f_\text{care} O_\text{c} \\
13.4 &\approx f_\text{gen}(O_\text{gd} + w + \bar F) + f_\text{dis}(O_\text{gd} + w + \bar F)
	+ f_\text{ess} (O_\text{e} + w + \bar F) + f_\text{care}(O_\text{c} + w + \bar F)\,.
\end{align*}

To close the system of equations and obtain our estimates, we require one further assumption.
 If $27\%$ of workers isolate at home, then the mean number of contacts at work is 
 \begin{equation*}
 	O_\text{gd}^* \approx 0.27 \times 0 + 0.73 \times (0.73 \,O_\text{gd}) \approx 0.5329 \,O_\text{gd}\,. 
\end{equation*}
We obtain $w \approx 5.14$, $w^* \approx 4$, $O_\text{e} \approx 16.23$, $O_\text{c} \approx 6.95$, $O_\text{gd} \approx 5.20$, and $O_\text{gd}^* \approx 3.08$. 
When we use {approximate} truncated power-law distributions to model the possibility that some people have many contacts and others have few contacts, we want to satisfy the following criteria:
\begin{itemize}
\item the general population has a mean of $w+O_\text{gd} \approx 10.34$ weak contacts per day when not physically distancing and a mean of $w^* + O_\text{gd}^* \approx 7.08$ weak contacts per day when physically distancing;
\item the disabled subpopulation has the same mean value of weak contacts as the general population whether or not people are physically distancing;
\item the caregiver subpopulation has a mean of $w \approx 5.14$ weak contacts per day when not physically distancing and a mean of $w \approx 4$ weak contacts per day when physically distancing; and
\item the essential-worker subpopulation has a mean of $O_\text{e} + w \approx 21.37$ weak contacts per day when not physically distancing and a mean of $O_\text{e} + w^* \approx 20.23$ contacts per day when physically distancing. 
\end{itemize}
Although the caregiver subpopulation may seem to have very few weak contacts, we note that most of their daily contacts come from $O_\text{c}$, which we estimate separately from the ordinary weak contacts.

Although the number of weak contacts for essential workers does decrease slightly during a lockdown, we use $O_\text{e} + w$ whether or not a lockdown is in place as an approximation because the difference in the numbers of weak contacts is very small ($21.37$ versus $20.23$). In practice, it was difficult for us to reduce the mean number of contacts slightly in this situation, because picking the minimum of two random variables of similar distributions tends to result in a value that is much smaller than the original one and doing so would result in the essential workers having far too few contacts.

\paragraph{Distribution of Strong Contacts:} We use data from
the 2016 Canadian census \cite{census} to describe the distribution of household sizes. 
According to these data, 105,750 households consist of 1 person, 124,280 households consist of 2 people, 58,010 households consist of 3 people, 55,215 households consist of 4 people, and 30,500 households consist of 5 or more people (which we treated as exactly 5 people). 
From these data, we construct an empirical distribution that we use for the entire population. It is $\mathcal{D}_s = \mathcal{E}(0.283, 0.332, 0.155, 0.148, 0.0816)$.

\paragraph{Caregivers:} To each disabled person, we assign one strong caregiver and one weak caregiver with whom they interact each day (although they do not interact with the latter when either they or the caregiver is symptomatic).
We chose the weak caregivers from a pool of caregivers. We use $10$ as the baseline caregiver-pool size, but we also consider other sizes ($4$ and $25$, as we discuss in Section \ref{sec:results}).

%%%%%

\subsection{Fits from Data} \label{fit-data}

There are three other parameters in our model that we also need to estimate. Even with our many estimates from the literature that we discussed in Section \ref{infer}, we still need to estimate the following quantities: (1) the maximum number $C^*$ of weak contacts that an individual has; (2) the number $A_0$ of people who are asymptomatic on day $0$; and (3) the probability $\tau$ that an individual who is symptomatically ill but not hospitalized is counted in the cumulative number of cases.

We model the number of weak contacts using an approximate truncated power-law distribution. That is, the daily number of weak contacts of an individual is distributed according to $\mathcal{P}(0, C^*; O_{q})$, where $O_{q}$ denotes the mean number of weak contacts of subpopulation $q$.

Based on the simulation procedure that we described in Section \ref{sec:simulate}, we use a fitting procedure (along with case data from Ottawa \cite{ottawa_case_counts})  to estimate 
$\tau$ and $C^*$ with a grid search. We use the first $90$ days as fitting data and assume that the associated contact distributions and mask-wearing policies are instantly adopted on day $44$ (i.e., the start of the lockdown in Ottawa). We tried fitting over shorter time windows, but these yielded poorer fits. The likely reason for the poor fits for these shorter time windows is that the parameter $C^*$ is smaller when fit over shorter time intervals (because the disease has spread less at that stage). The longer time window allows $C^*$, which may be a key driver in the disease dynamics, to be fit to a larger value and thereby allow extensive spreading of the disease.

We assume that there are $A_0$ people on day $0$ in the asymptomatic compartment and that all other individuals are in the susceptible compartment. On day $1$, with the first recorded case, there is $1$ recorded case in expectation. 
Therefore, 
\begin{equation}
	1 = \underbrace{\frac{\alpha}{\alpha+\eta}}_{{\Pr(\text{transition from asymptomatic to ill})}} \times \underbrace{\e^{-(\alpha+\eta) \times 1 \text{day}}}_{\Pr(\text{leave the A compartment in 1 day})} \times \underbrace{\tau}_{\text{\ \ test ill individual}} \times A_0\,. \label{eq:A0}
\end{equation}	
The first factor is the probability that the transition from the A compartment to the I compartment occurs before the transition from A to the R compartment. The second factor is the probability there is a transition out of the A compartment in a $1$-day time period. The third factor is the probability that an individual in the I compartment tests positive for COVID-19. The fourth factor ($A_0$) is the total number of asymptomatic people on day $0$.} Our choice to make the expected number of documented cases equal to $1$ on day $1$ allows us to have two parameters (rather than three) when fitting. Using more parameters can result in overfitting. 

We seek to minimize the $\ell_2$-error in new daily cases (i.e., the change in the daily cumulative case count). Because our stochastic model is complicated, with variation across trials, we use a grid search (instead of a gradient-based method) to estimate parameters. In Table \ref{tab:fit_errs}, we summarize our results. From this procedure, our ``optimal'' parameter values are $\tau = 0.04$ and $C^* = 60$.

\begin{table}
	\begin{center}
		\begin{tabular}{c | c | c}
			$\mathbf{\tau}$ & $\mathbf{C^*}$ & \bf Error \\
			\hline
$0.02$ & $50$ & $2.31 \times 10^4$\\
$0.03$ & $50$ & $1.82 \times 10^4$\\
$0.03$ & $60$ & $2.38 \times 10^4$\\
$0.04$ & $50$ & $2.13 \times 10^4$\\
${\bf 0.04}$ & ${\bf 60}$ & \bf ${\bf 1.74} \times {\bf 10}^{\bf 4}$\\
$0.04$ & $70$ & $3.05 \times 10^4$\\
$0.05$ & $50$ & $2.48 \times 10^4$\\
$0.05$ & $60$ & $1.83 \times 10^4$\\
$0.05$ & $70$ & $2.26 \times 10^4$\\
$0.06$ & $50$ & $2.80 \times 10^4$\\
$0.06$ & $60$ & $2.01 \times 10^4$\\
$0.06$ & $70$ & $1.76 \times 10^4$\\
$0.06$ & $80$ & $3.48 \times 10^4$\\
$0.07$ & $50$ & $3.16 \times 10^4$\\
$0.07$ & $60$ & $2.26 \times 10^4$\\
$0.07$ & $70$ & $1.80 \times 10^4$\\
$0.07$ & $80$ & $2.78 \times 10^4$\\
$0.08$ & $50$ & $3.32 \times 10^4$\\
$0.08$ & $60$ & $2.38 \times 10^4$\\
$0.08$ & $70$ & $1.86 \times 10^4$\\
$0.08$ & $80$ & $2.51 \times 10^4$\\
$0.09$ & $50$ & $3.55 \times 10^4$\\
$0.09$ & $60$ & $2.69 \times 10^4$\\
$0.09$ & $70$ & $1.93 \times 10^4$\\
$0.09$ & $80$ & $2.25 \times 10^4$\\
$0.10$ & $50$ & $3.61 \times 10^4$\\
$0.10$ & $60$ & $2.82 \times 10^4$\\
$0.10$ & $70$ & $2.07 \times 10^4$\\
$0.10$ & $80$ & $1.96 \times 10^4$\\
$0.10$ & $90$ & $4.40 \times 10^4$\\
$0.11$ & $50$ & $3.80 \times 10^4$ \\
$0.11$ & $60$ & $2.88 \times 10^4$ \\
$0.11$ & $70$ & $2.16 \times 10^4$ \\
$0.11$ & $80$ & $1.99 \times 10^4$ \\
$0.11$ & $90$ & $3.54 \times 10^4$ \\
			\hline			
		\end{tabular} \caption{The $\ell_2$-error in new daily documented cases
		for various values of $\tau$ and $C^*$. 		
		Using \eqref{eq:A0}, with values of $\alpha$ and $\mu$ from the literature and a given value of $\tau$, we determine $A_0$. For each set of parameters, we conduct $96$ trials and we compute the error by taking the mean of all trials in which there are at least $250$ documented cases {through} day $90$. We only report parameter values for which the errors are smaller than $5 \times 10^4$. We test all parameter values on the lattice $(\tau, C^*) \in \{0.01, 0.02, 0.03, 0.04, 0.05, 0.06, 0.07, 0.08, 0.09. 0.10, 0.11 \} \times \{ 50, 60, 70, 80, 90, 100, 110, 120, 130, 140, 150 \}$. We show our best results in bold. That is, our ``optimal'' parameter values (see the sixth row) are $\tau = 0.04$ and $C^* = 60$.}
\label{tab:fit_errs}
	\end{center}
\end{table}

%%%%

\section{Simulations of our Stochastic Model of COVID-19 Spread}
\label{sec:simulate}

%%%%

\subsection{Simulation Steps}

We summarize our simulation procedure in Algorithm \ref{alg:overall}. Note that it uses the other algorithms that we present in this subsection. The code is available at our \href{https://3k1m@bitbucket.org/3k1m/covid19-disabledcaregiverstudy.git}{Bitbucket repository.}

We initially construct a network by matching ends of edges (i.e., ``stubs'') in a generalization of a configuration-model network. For weak contacts, we assign a number of stubs to each individual in each subpopulation to encode their number of weak contacts (see Algorithm \ref{alg:weakstubs}). We determine this number from an associated probability distribution. We then do a so-called ``random matching'' (see Algorithm \ref{alg:weak}), in which we match stubs uniformly at random. Pairs of individuals whose stubs are matched are contacts of each other. If we choose two individuals who are already contacts or an individual is paired with themself, we simply discard that pairing. For strong contacts, we assign individuals to units (see Algorithm \ref{alg:unit}) and make members of these units strong contacts with each other unless they are already contacts (see Algorithm \ref{alg:strong}). Consequently, the number of contacts per individual does not perfectly match the desired distributions. However, for a network with many nodes, these errors are negligible in practice. See \cite{fosdick2018} for a detailed exposition of different types of configuration models (although we employ a generalization of a configuration model), including different strategies for how to deal with self-edges and multi-edges. We assign weak and strong caregivers to disabled people in a manner that is analogous to how we assign strong contacts (see Algorithm \ref{alg:disabled}).

We then place some number of individuals, who we choose uniformly at random from the nodes in the network, into the A and/or I compartments. This number of individuals, which subpopulation they belong, and the choice of these compartments (all of these individuals in A, all of these individuals in I, or some of these individuals in A and some of them in I) depends on user input. For example, in the four simulations that we used to generate Figure \ref{fig:Seeding}, all initially infected individuals are in the A compartment and belong to the general subpopulation, caregiver subpopulation,  disabled subpopulation, and essential-worker subpopulation, respectively. In all other simulations that we discuss in the present paper, the initially infected individuals are all in the A compartment. After having initialized the contact structures, we execute the commands in the following paragraphs for a user-specified number of iterations.

We check if we need to update contact structures and/or mask-wearing strategies because of a lockdown (see Algorithm \ref{alg:close}) or a reopening (see Algorithm \ref{alg:open}). For a lockdown, we update the mask-wearing strategies and assign each individual a number of weak contacts from the new weak-contact distributions. If the new number of weak contacts is less than the current number of weak contacts, we remove excess contacts uniformly at random. For a reopening, we again update the mask-wearing strategies and assign each individual a number of weak contacts from the new weak-contact distribution. If the new 
number of weak contacts is larger than the current number of weak contacts, we assign the individual a number of stubs that is equal to the difference between the new sample and the current number of contacts and apply Algorithm \ref{alg:weak} to connect the stubs.

On each day, we assign a weak caregiver to each disabled person uniformly at random from their pool of weak caregivers, as long as neither is breaking their contacts. We then use Algorithm \ref{alg:advance} to determine if each individual in the network remains in their current compartment or moves to a new one. If an individual is in the S compartment, we calculate the probability of infection using Algorithm \ref{alg:infect}. In this algorithm, we loop through each of this individual's contagious contacts (i.e., those in the {A, I, or H} compartments) and use Equation \ref{eq:notget} to calculate the probability that the individual becomes infected. For compartments E and H, for which there is only one possible transition to a new compartment, we draw a transition time from $\text{Exp}(\chi)$ (where $\chi$ is the associated rate constant) to determine if there is a transition between compartments. If the time is less than $1$ day, then the individual changes compartments; otherwise, the individual stays in their current compartment. For compartments A and I, 
from which an individual can move to one of two possible new compartments, we draw transition times from $\text{Exp}(\chi_1)$ and $\text{Exp}(\chi_2)$, where $\chi_1$ and $\chi_2$ are the associated rate constants. If both times are less than $1$ day, the individual moves to the compartment that has the smaller time. If only one of the times is less than $1$ day, the individual moves to that compartment. If neither time is less than $1$ day, the individual remains in their current compartment. When an individual moves to the I compartment, they may break their weak contacts. With probability $b$, they break all of their weak contacts; otherwise, they keep all of their weak contacts. Individuals in the I compartment become documented cases with probability $\tau$. In our pseudocode, we refer to the breaking of contacts as ``deactivating'' edges and refer to the re-establishment of contacts as ``reactivating'' edges. If an individual moves to the H compartment, we deactivate all of their edges with weak and strong contacts. If an individual moves to the R compartment, we reactivate any edges that may have been deactivated because of their movement through the I and H compartments (except those that may not be active because (1) the other individual in the interaction is in the I compartment and did not break their weak connection or (2) the other individual is in the H compartment).

%%%%%

\begin{algorithm}
\caption{A Simulation of the Spread of COVID-19 on a Contact Network}
\label{alg:overall}
\hspace*{\algorithmicindent} \textbf{Input:} {A set of values for each parameter that we list in Table \ref{tab:params}} \\
\hspace*{\algorithmicindent} \textbf{Output:} Daily counts of people in each compartment; documented cases
\begin{algorithmic}[1]
\STATE{Initialize {\it Population} of size {\it P}\textsubscript{Ottawa} with fractions {\it f}\textsubscript{dis} who are disabled, {\it f}\textsubscript{care} who are caregivers, {\it f}\textsubscript{ess} who are essential workers, and {\it f}\textsubscript{gen} who are members of the general population. At initialization, we determine whether or not each individual will break all of their weak contacts if they become ill (they break weak contacts with probability $b$) and determine whether or not they will have a positive test result if they become ill (a positive test occurs with probability $\tau$).}
\STATE{Assign a unique integer ID to each individual in {\it Population}.}
\STATE{Obtain {\it WeakStubs} from Algorithm \ref{alg:weakstubs} with input {\it Population}.}
\STATE{{Obtain {\it PossibleHouseholdUnits} from Algorithm \ref{alg:unit} with input {\it Population}.}}
\STATE{{Assign weak contacts using Algorithm \ref{alg:weak} with inputs {\it Population, WeakStubs}.}}
\STATE{{Assign strong contacts using Algorithm \ref{alg:strong} with inputs {\it Population, PossibleHouseholdUnits}.}}
\STATE{{Match disabled people and caregivers using Algorithm \ref{alg:disabled} with input {\it DisabledPopulation}, where {\it DisabledPopulation} refers to all individuals in {\it Population} who are in the disabled subpopulation.}}
\STATE{Initialize some number of people to be asymptomatic or ill based on program inputs.}
(In all of our simulations in the present paper, we initialize these individuals to be asymptomatic, but one can instead use our code to initialize individuals as ill; one can also initialize some individuals to be asymptomatic and some individuals to be ill.)
\STATE{{\it day }$=0$, {\it has\_opened} $=$ \FALSE, {\it has\_closed} $=$ \FALSE}
\WHILE{{\it day} $<$ {\it end\_day}}
\STATE{Compute the number of individuals from each subpopulation in each compartment, as well as the number of documented cases from each subpopulation.}
\FORALL{{\it disabled\_individual} in {\it DisabledPopulation}}
\STATE{Select a weak caregiver uniformly at random from their set of weak caregivers.}
\ENDFOR
\FORALL{{\it individual} in {\it Population}} 
\STATE{Calculate the infection probability using Algorithm \ref{alg:infect} with input {\it individual}.}
\ENDFOR
\FORALL{{\it individual} in {\it Population}}
\STATE{Advance state by $1$ day using Algorithm \ref{alg:advance} with input {\it individual}.}
\ENDFOR
\STATE{{\it day } $\gets \text{{\it day}} + 1$}
\IF{{\it time} $<$ {\it close\_time}}
\STATE{Do nothing.}
\ELSIF{{\it time} $<$ {\it open\_time}}
\IF {\NOT {\it has\_closed}}
\STATE{Close down (i.e., start a lockdown) using Algorithm \ref{alg:close}.}
\STATE{{\it has\_closed} $\gets$ \TRUE}
\ENDIF
\ELSE
\IF {\NOT {\it has\_opened}}
\STATE{Reopen (i.e., end a lockdown) using Algorithm \ref{alg:open}.} 
\STATE{{\it has\_opened} $\gets$ \TRUE}
\ENDIF
\ENDIF
\ENDWHILE
\end{algorithmic}
\end{algorithm}

\begin{algorithm}
\caption{Weak Stubs}
\label{alg:weakstubs}
\hspace*{\algorithmicindent} \textbf{Input:} A container of nodes (which we denote by {\it Population})\\
\hspace*{\algorithmicindent} \textbf{Output:} A container of IDs (which we denote by {\it WeakStubs}) in which the ID of each node in {\it Population} occurs  \\
\hspace*{\algorithmicindent} with a multiplicity that is equal to the number of stubs of that node.
\begin{algorithmic}[1]
\FORALL{{\it individual} in {\it Population}}
\STATE{Let {\it target} equal the number of weak stubs that {\it individual} can potentially have; we draw this number from $\mathcal{D}_{\text{group,period}}$, where ``group'' is their subpopulation and ``period'' is the current state of the pandemic (pre-lockdown, lockdown, or post-lockdown).}
\STATE{Let {\it current} equal the number of current weak stubs of {\it individual}}.
\IF{{\it current} $<$ {\it target}}
\STATE{$\text{{\it needed}} = \text{{\it current}} - \text{{\it target}}$}
\ELSE
\STATE{{\it needed} $= 0$}
\ENDIF
\STATE{For {\it needed} number of times, append the ID of {\it individual} to a container {\it WeakStubs}}.
\RETURN{{\it WeakStubs}}
\ENDFOR
\end{algorithmic}
\end{algorithm}

\begin{algorithm}
\caption{Household Units}
\label{alg:unit}
\hspace*{\algorithmicindent} \textbf{Input:} A container of nodes (which we denote by {\it Population}) \\
\hspace*{\algorithmicindent} \textbf{Output:} A container of containers of IDs (which we denote by {\it PossibleHouseholdUnits})
\begin{algorithmic}[1]
\STATE{Let {\it AllIDs} be a container that stores the unique ID for each {individual} in {\it Population}} 
\WHILE{{\it AllIDs} \NOT empty}
\STATE{Choose an ID, which we denote by {\it ID\textsubscript{1}}, uniformly at random from {\it AllIDs} and determine the number of household contacts ({\it house}) of the individual with that ID by sampling from $\mathcal{D}_\text{strong}.$}
\STATE{Select {\it house} number of IDs uniformly at random from {\it AllIDs}.}
\STATE{{Append {\it ID\textsubscript{1}} and the above IDs to a container {\it unit.}}}
\STATE{Remove {all of the IDs in {\it unit} from {\it AllIDs}}.}
\ENDWHILE
\RETURN{{\it PossibleHouseholdUnits} (which is a container that holds each {\it unit})}
\end{algorithmic}
\end{algorithm}

\begin{algorithm}
\caption{Assigning Weak Contacts}
\label{alg:weak}
\raggedright
\hspace*{\algorithmicindent} \textbf{Input:} A container of nodes (which we denote by {\it Population}) and a container of IDs (which we denote by 
\hspace*{\algorithmicindent} {\it WeakStubs})\\
\hspace*{\algorithmicindent} \textbf{Result:} All nodes in {\it Population} are assigned weak contacts
\begin{algorithmic}[1]
\WHILE{$|\text{{\it WeakStubs}}| \geq 2$}
\STATE{Choose IDs {\it ID\textsubscript{1}} and {{\it ID\textsubscript{2}}} uniformly at random from {\it WeakStubs}.}
\IF{{{\it ID\textsubscript{1}}} $\neq$ {{\it ID\textsubscript{2}}} \AND {the individuals with IDs} {{\it ID\textsubscript{1}}} and {{\it ID\textsubscript{2}}} are not already contacts (weak, strong, or caregiving)}
\STATE{Make {the individuals with IDs} {{\it ID\textsubscript{1}}} and {{\it ID\textsubscript{2}}} into weak contacts.}
\ENDIF
\STATE{Remove {{\it ID\textsubscript{1}}} and {{\it ID\textsubscript{2}}} from {\it WeakStubs}.}
\ENDWHILE 
\end{algorithmic}
\end{algorithm}

\begin{algorithm}
\caption{Assigning Strong Contacts}
\label{alg:strong}
\hspace*{\algorithmicindent} \textbf{Input:} A container of nodes (which we denote by {\it Population}) and a container of containers of IDs (which we denote \\
\hspace*{\algorithmicindent} by {\it PossibleHouseholdUnits}) \\
\hspace*{\algorithmicindent} \textbf{Result:} {All nodes in {\it Population} are assigned strong contacts} 
\begin{algorithmic}[1]
\FORALL{{\it unit} in {\it PossibleHouseholdUnits}}
\FORALL{{\it ID} in {\it unit}}
\STATE{Make the individual with {\it ID} a strong contact of each other member of the unit, unless the individuals are already contacts (weak, strong, or caregiving).} 
\ENDFOR
\ENDFOR
\end{algorithmic}
\end{algorithm}

\begin{algorithm}
\caption{Matching Disabled People and Caregivers}
\label{alg:disabled}
\raggedright
\hspace*{\algorithmicindent} \textbf{Input:} A container of nodes (which we denote by {\it DisabledPopultion}) that all belong to the same subpopulation \\
\hspace*{\algorithmicindent} {\textbf{Result:} All nodes in {\it DisabledPopulation} are assigned a pool of weak caregivers and one strong caregiver}\\
\begin{algorithmic}[1]
\FORALL{{\it disabled\_individual} in {\it DisabledPopulation}}
\STATE{Determine {\it care\_weak\_num} from $\mathcal{D}_\text{pool}$, which is the number of weak caregivers in their pool.}
\STATE{Select {\it care\_weak\_num} number of caregivers uniformly at random from the set of caregivers and store them in {\it CaregiversChosen}}.
\FORALL{{\it caregiver} in {\it CaregiversChosen}}
\IF{{\it disabled\_individual} and {\it caregiver} are not already contacts (weak, strong, or caregiving)}
\STATE{Make their relationship a weak caregiver--disabled relationship.}
\ENDIF
\ENDFOR
\ENDFOR
\FORALL{{\it disabled\_individual} in {\it DisabledPopulation}}     
\STATE{Choose $1$ caregiver uniformly at random from the set of caregivers.}
\IF{{\it disabled\_individual} and the caregiver are not already contacts (weak, strong, or caregiving)}
\STATE{Make their relationship a strong caregiver--disabled relationship.}
\ENDIF
\ENDFOR
\end{algorithmic}
\end{algorithm}

\begin{algorithm}
\caption{{Infection Probability}}
\label{alg:infect}
\hspace*{\algorithmicindent} \textbf{Input:} A node (which we denote by {\it individual}) \\
\hspace*{\algorithmicindent} \textbf{Output:} An infection probability (which we denote by {\it infect\_prob}) 
\begin{algorithmic}[1]
\STATE{{\it not\_get} $=1$}
\IF{{\it individual} is Susceptible}
\FORALL{{\it weak\_contact} in {{\it individual}'s weak contacts}} 
\IF{edge to {\it weak\_contact} is {active}
\AND {\it weak\_contact} is contagious}
\IF{both wear a mask}
\STATE{{\it not\_get} {$\gets \text{{\it not\_get}} \,\times (1 - \beta m w_\text{w})$}}
\ELSE
\STATE{{\it not\_get} 
$\gets \text{{\it not\_get}} \,\times (1 - \beta w_\text{w})$}
\ENDIF
\ENDIF
\ENDFOR
\FORALL{{\it strong\_contact} in {\it individual}'s strong contacts}
\IF{edge to {\it strong\_contact} is active \AND {\it strong\_contact} is contagious}
\STATE{{\it not\_get} $ \gets \text{{\it not\_get}} \,\times (1 - \beta w_\text{s})$}
\ENDIF
\ENDFOR
\IF{{\it individual} is disabled}
\FORALL{{\it caregiver} in their set of weak caregivers for the day}
\IF{edge to {\it caregiver} is active and {\it caregiver} is contagious}
\IF{both wear a mask}
\STATE{{\it not\_get} $\gets \text{{\it not\_get}} \,\times (1 - \beta m w_\text{c})$}
\ELSIF{one wears a mask}
\STATE{{\it not\_get} $\gets \text{{\it not\_get}} \,\times (1 - \beta \sqrt{m} w_\text{c})$}
\ELSE
\STATE{{\it not\_get} $\gets \text{{\it not\_get}} \,\times (1 - \beta w_\text{c})$}
\ENDIF
\ENDIF
\ENDFOR
\IF{edge to {\it individual}'s strong caregiver is active and the strong caregiver is contagious}
\IF{both wear a mask}
\STATE{{\it not\_get} $\gets \text{{\it not\_get}} \,\times (1 - \beta m w_\text{c})$}
\ELSIF{one wears mask}
\STATE{{\it not\_get} $\gets \text{{\it not\_get}} \,\times (1 - \beta \sqrt{m} w_\text{c})$}
\ELSE
\STATE{{\it not\_get} $\gets \text{{\it not\_get}} \,\times (1 - \beta w_\text{c})$}
\ENDIF
\ENDIF
\ELSIF{{\it individual} is a caregiver}
\FORALL{{\it disabled\_individual} in their set of disabled contacts for the day}
\IF{edge to {\it disabled\_individual} is active and {\it disabled\_individual} is contagious}
\IF{both wear a mask}
\STATE{{\it not\_get} $\gets \text{{\it not\_get}} \,\times (1 - \beta m w_\text{c})$}
\ELSIF{one wears mask}
\STATE{{\it not\_get} $\gets \text{{\it not\_get}} \,\times (1 - \beta \sqrt{m} w_\text{c})$}
\ELSE
\STATE{{\it not\_get} $\gets \text{{\it not\_get}} \,\times (1 - \beta w_\text{c})$}
\ENDIF
\ENDIF
\ENDFOR
\ENDIF
\ENDIF
\STATE{{\it infect\_prob} $= 1 - \text{{\it not\_get}}$}
\RETURN{{{\it infect\_prob}}}
\end{algorithmic}
\end{algorithm}

\begin{algorithm}
\caption{Advancing One Day}
\label{alg:advance}
\hspace*{\algorithmicindent} \textbf{Input:} A node (which we denote by {\it individual}) \\
\hspace*{\algorithmicindent} {\textbf{Result:} {\it individual} remains in their current compartment or moves to a new one} 
\begin{algorithmic}[1]
\IF{{\it individual} is in the S compartment}
\STATE{In the time interval $\Delta T = 1$ day, move {\it individual} into the E compartment with probability {\it infect\_prob.}}
\ELSIF{{\it individual} is in the E compartment}
\STATE{Sample $T_\text{asymptomatic}$ from the distribution $\text{Exp}(\nu)$.}
\IF{$T_\text{asymptomatic} < 1$ day}
\STATE{Move {\it individual} to the A compartment.}
\ENDIF
\ELSIF{{\it individual} is in the A compartment}
\STATE{Sample $T_\text{ill}$ from $\text{Exp}(\alpha)$.}
\STATE{Sample $T_\text{removed}$ from $\text{Exp}(\eta)$.}
\IF{$T_\text{ill} < T_\text{removed}$}
\IF{$T_\text{ill}< 1$ day}
\STATE{Move {\it individual} to the I compartment.}
\STATE{Deactivate all of their edges to weak contacts if that ill individual is one who breaks their weak contacts.}
\ENDIF
\ELSE 
\IF{$T_\text{removed} < 1$ day}
\STATE{Move {\it individual} to the R compartment.}
\STATE{Reactivate all of their edges to weak contacts {(provided either that the weak contact has no symptoms or that the weak contact is in the I compartment but does not break weak contacts when ill).}} 
\ENDIF
\ENDIF 
\ELSIF{{\it individual} is in the I compartment}
\STATE{Sample $T_\text{hospital}$ from $\text{Exp}(\mu)$.}
\STATE{Sample $T_\text{removed}$ from $\text{Exp}(\rho)$.}
\IF{$T_\text{hospital} < T_\text{removed}$}
\IF{$T_\text{hospital}< 1$ day}
\STATE{Move {\it individual} to the H compartment.}
\STATE{Deactivate all of their edges to weak and strong contacts.}
\ENDIF
\ELSE 
\IF{$T_\text{removed} < 1$ day}
\STATE{Move {\it individual} to the R compartment.}
\STATE{Reactivate all of their edges to weak contacts {(provided either that the weak contact has no symptoms or that the weak contact is in the I compartment but does not break weak contacts when ill).}}  
\ENDIF
\ENDIF
\ELSIF{{\it individual} is in the H compartment}
\STATE{Sample $T_\text{removed}$ from $\text{Exp}(\zeta)$.}
\IF{$T_\text{removed} < 1$ day}
\STATE{Move {\it individual} to to the R compartment.}
\STATE{Reactivate all of their edges to weak contacts {(provided either that the weak contact has no symptoms or that the weak contact is in the I compartment but does not break weak contacts when ill).}}
\ENDIF
\ENDIF
\end{algorithmic}
\end{algorithm}

\begin{algorithm}
\caption{Closing Down (i.e., starting a lockdown)}
\label{alg:close}
\hspace*{\algorithmicindent} \textbf{Input:} A container of nodes (which we denote by {\it Population}) \\
\hspace*{\algorithmicindent} \textbf{Result:} Lockdown mask-wearing strategies and contact-limiting strategies are implemented for each node in {\it Population}
\begin{algorithmic}[1]
\STATE{Update mask-wearing statuses.}
\FORALL{{\it individual} in {\it Population}}
\STATE{{Determine their new number of weak contacts by sampling {\it new\_target\_value}} from $\mathcal{D}_\text{group,post}$, where {\it group} is the subpopulation of the individual.}
\ENDFOR
\FORALL{{\it individual} 
in {\it Population}}
\STATE{$\text{{\it clear}} = \max\{0, \text{{\it current\_weak\_contacts}} - \text{{\it new\_target\_value\}}}$}
\STATE{$i = 0$}
\WHILE{$i <$ \textit{clear}}
\STATE{Select a weak contact $\varpi$ uniformly at random.} 
\IF{{neither $\varpi$ nor {\it individual} is an essential worker}} 
\STATE{Remove the edge between the nodes.}
\ENDIF
\STATE{$i \gets i +1$}
\ENDWHILE
\ENDFOR
\end{algorithmic}
\end{algorithm}

\begin{algorithm}
\caption{Reopening (i.e., ending a lockdown)}
\label{alg:open}
\raggedright
\hspace*{\algorithmicindent} \textbf{Input:} A container of nodes (which we denote by {\it Population}) \\
\hspace*{\algorithmicindent} {\textbf{Output:}  Reopening mask-wearing strategies and contact-limiting strategies are implemented for each node in 
\hspace*{\algorithmicindent} {\it Population}}
\begin{algorithmic}[1]
\STATE{Update mask-wearing statuses.}
\STATE{{Obtain a container {\it new\_weak\_stubs} by applying Algorithm \ref{alg:weakstubs} with input {\it Population}}}
\STATE{Apply Algorithm \ref{alg:weak} with inputs {\it Population} and {\it new\_weak\_stubs}.}
\end{algorithmic}
\end{algorithm}

%%%%

\subsection{Implementation of Approximate Truncated Power-Law Distributions}

%%%

\subsubsection{Sampling from the Distribution}

Given a lower bound $a_-$, an upper bound $a_+$, and an exponent $p$, we wish to approximate a power-law distribution for a discrete random variable $N$ over $[a_-,a_+]$, where $\Pr(N=n) = O(n^{-p})$ as $a_+, n \rightarrow \infty$. Our procedure amounts to (1) shifting the range to avoid the case $a_-=0$, (2) sampling from a continuous power-law probability density, (3) truncating the result to an integer, and (4) shifting the range back if we shifted the original range away from $a_-=0$. In our model, we use $a_-=0$ and $a_+ = C^*$, but we present the approach for a general finite sequence of nonnegative integers.

If $a_-=0$, we first shift to a distribution on $[A,B]$, where $A=\max\{a_-,1\}$ and $B=a_++(A-a_-)$. We define the normalization constant 
\begin{align} 
	C &= \int_A^{B+1} x^{-p} \text{d} x \nonumber \\
&= \begin{cases} \frac{1}{1-p} ( (B+1)^{1-p} - A^{1-p})\,, \quad p \neq 1 \\
	\log(\frac{B+1}{A})\,, \quad p = 1\,. \end{cases} \label{eq:Cnorm}
\end{align}
To choose $N$, we select $u \in [0,1)$ from a uniform distribution and select $x^*$ such that 
\begin{equation} C^{-1} \int_A^{x^*} x^{-p} \text{d}x = u. \end{equation}  We then calculate 
\begin{equation} n^* = \lfloor x^* \rfloor\,, \label{eq:nstar} \end{equation} where $\lfloor z \rfloor$ is the floor of $z$ (i.e., the largest integer that is less than or equal to $z$). That is,
\begin{equation}
	x^* = \begin{cases} ((1-p) uC + A^{1-p})^{1/(1-p)}\,, \quad p \neq 1 \\
A \exp(uC)\,, \quad p = 1\,. \end{cases} \label{eq:xstar}
\end{equation}
Finally, we shift back to set \begin{equation} N = n^* - (A-a_-)\end{equation}.

Note that 
\begin{align*} 
	\Pr(N=n) &\propto \int_{n+(A-a_-)}^{n + 1 + (A-a_-)} x^{-p} \dd x = \begin{cases} \log(\frac{ n + 1 + (A-a_-)}{n + (A-a_-)})\,, \quad p = 1 \\
\left|(n+1+(A-a_-))^{1-p} - (n+(A-a_-))^{1-p}\right|\,, \quad p \neq 1 \end{cases} \\
	&= \begin{cases} \log(1 + \frac{1 }{n + (A-a_-)} )\,, \quad p = 1 \\
\left|(n+(A-a_-))^{1-p} \left( 1 + \frac{1}{N + (A-a_-)}  \right)^{1-p} - 1\right|\,, \quad p \neq 1 \end{cases} \\
	&= O(1/n^p) \quad \text{as} \quad n \rightarrow \infty\,, 
\end{align*} 
thereby ensuring that asymptotically we have a power law as $n \rightarrow \infty$.

%%%%%

\subsubsection{Estimating the Mean}

When $B-A$ is large, it can be computationally expensive to compute the precise mean of the random variable $N = n^* - (A-a_-)$  that we obtain from {Eqs.~\eqref{eq:Cnorm}--\eqref{eq:xstar}}. When $B-A$ is large, it is also the case that rounding errors and overflow errors can cause an estimation of the true mean to be inaccurate. Therefore, we estimate the mean analytically. Given $a_-$, $a_+$, and $p$, we seek to estimate the mean $E_p := \mathbb{E}(N)$ over the {interval $[a_-,a_+]$.}

We have that 
\begin{align*} 
	E_p &= C^{-1} \sum_{n=A}^B n \int_{n}^{n+1} x^{-p} \dd x \\
		&= \begin{cases} C^{-1} \sum_{n=A}^b n \log((n+1)/n)\,, \quad p=1 \\
((1-p)C)^{-1} \sum_{n=A}^b n ( (n+1)^{1-p} - n^{1-p})\,, \quad p \neq 1 \end{cases} \\
		&= \begin{cases} C^{-1} \sum_{n=A}^b n \log((n+1)/n)\,, \quad p=1 \\
	((1-p)C)^{-1} \left( (B+1)^{2-p} - A^{2-p} - \sum_{n=A+1}^{B+1} n^{1-p} \right), \quad p \neq 1\,. \end{cases}
 \end{align*} 
To obtain the third equality, we rewrote $\sum_{n=A}^B n(n+1)^{1-p}$ as $\sum_{n=A+1}^{B+1} (n-1)n^{1-p}$, whose $n^{2-p}$ terms cancel with $\sum_{n=A}^B n^{2-p}$ except at $n=A$ and $n=B+1$.

For our approximation, we consider multiple cases.

\paragraph{$p=1$:} Note that $\{n \log(\frac{n+1}{n})\}_{n=A}^{B}$ is an {increasing sequence} of terms. Therefore, 
\begin{equation*}
	\underline{S}_1 := A \log\left(\frac{A+1}{A}\right) + \int_A^B x \log\left(\frac{x+1}{x}\right) \dd x \leq \sum_{n=A}^B n \log\left(\frac{n+1}{n}\right) \leq \int_A^{B+1} x \log\left(\frac{x+1}{x}\right) \dd x =: \overline{S}_1\,. 
\end{equation*}	
Because $\int x \log(\frac{x+1}{x}) \dd x = \frac{1}{2} ( x^2 \log((x+1)/x) + x - \log(x+1)) + \text{const}$, we compute the integrals exactly and obtain the estimate $E_1 = C^{-1} (\underline{S}_1 + \overline{S}_1)/2$.

\paragraph{$p=2$:} We need to estimate $\sum_{n=A+1}^{B+1} n^{-1}$. Because the sequence $1/n$ is decreasing, 
\begin{equation*}
	\underline{S}_2 := \int_{A+1}^{B+2} \frac{1}{x} \dd x = \log\left(\frac{B+2}{A+1}\right) \leq \sum_{n=A+1}^{B+1} n^{-1} \leq \frac{1}{A+1} + \log\left(\frac{B+1}{A+1}\right) = \frac{1}{A+1} + \int_{A+1}^{B+1} \frac{1}{x} \dd x =: \overline{S}_2\,.
\end{equation*}
We then estimate $E_2 = C^{-1} \left( \frac{1}{2}(\underline{S}_2 + \overline{S}_2 ) \right)$, where $A^{2-p} = (B+1)^{2-p} = 1$ allows us to cancel terms.

\paragraph{$p \notin \{1,2\}\,,\, p > 1$:} We need to estimate $\sum_{n=A+1}^{B+1} n^{1-p}$, where the terms are decreasing. Therefore,
\begin{align*} 
	\underline{S}_{p>} &:= \int_{A+1}^{B+2} x^{1-p} \dd x = \frac{(B+2)^{2-p} - (A+1)^{2-p}}{2-p} \leq \sum_{n=A+1}^{B+1} n^{1-p} \\ &\leq (A+1)^{1-p} + 			\frac{(B+1)^{2-p} - (A+1)^{2-p}}{2-p} =  (A+1)^{1-p} + \int_{A+1}^{B+1} x^{1-p} \dd x =: \overline{S}_{p>}\,. 
\end{align*} 
We then estimate $E_{p>} = ((1-p)C)^{-1} \left( (B+1)^{2-p} - A^{2-p} - \frac{1}{2}(\underline{S}_{p>} + \overline{S}_{p>} ) \right).$

\paragraph{$p \notin \{1,2\}\,,\, p < 1$:} We need to estimate $\sum_{n=A+1}^{B+1} n^{1-p}$, where the terms are increasing. Therefore,
\begin{align*} 
	\underline{S}_{p<} &:= (A+1)^{1-p} + \int_{A+1}^{B+1} x^{1-p} \dd x = (A+1)^{1-p} + \frac{(B+1)^{2-p} - (A+1)^{2-p}}{2-p} \leq \sum_{n=A+1}^{B+1} n^{1-p} \\ &\leq \frac{(B+2)^{2-p} - (A+1)^{2-p}}{2-p} = \int_{A+1}^{B+2} x^{1-p} \dd x =: \overline{S}_{p<}\,. 
\end{align*} 
We then estimate $E_{p<} = ((1-p)C)^{-1} \left( (B+1)^{2-p} - A^{2-p} - \frac{1}{2}(\underline{S}_{p<} + \overline{S}_{p<} ) \right).$

\begin{figure}
\centering
\includegraphics[width=5.in]{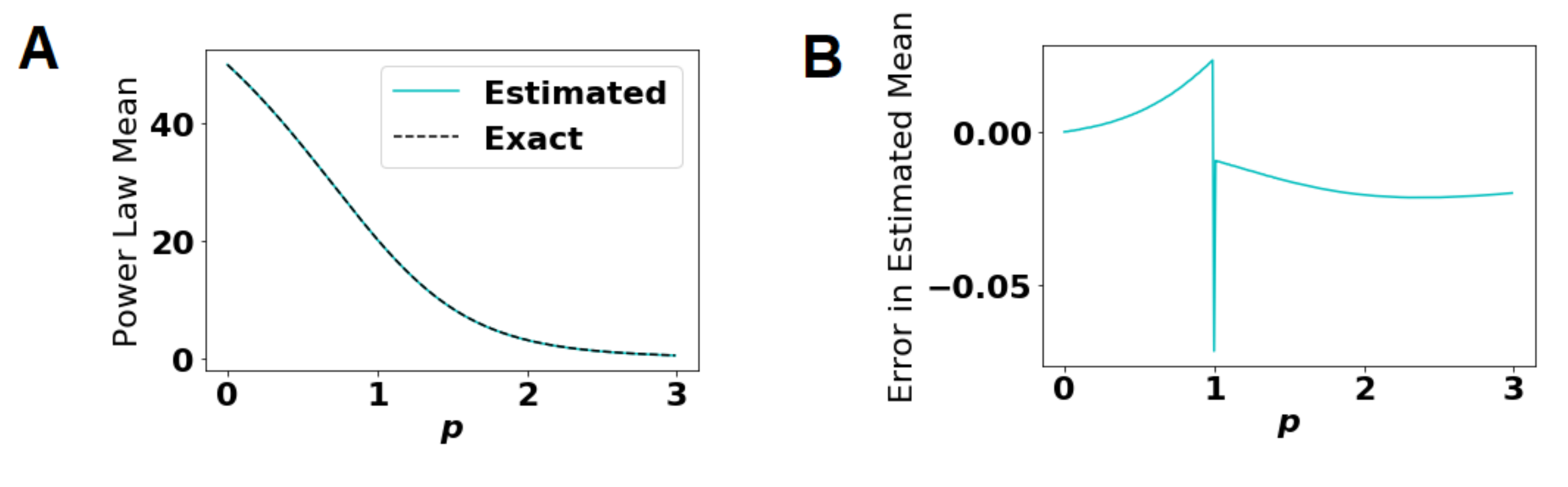}
\caption{(A) The estimated and exact mean values {of the approximate truncated power-law distribution} for various values of $p$. The curves are indistinguishable. (B) The {error in computing the mean for our approximations}. In this figure (both panel A and panel B), we use $a_-=0$ and $a_+=100$. 
} \label{fig:mean_estimates}
\end{figure}

This approximation is very accurate. When $a_-=0$ and $a_+=100$, we plot the approximations and the numerically exact values in Fig.~\ref{fig:mean_estimates}.

%%%%%

\section{Additional Computational Experiments}

\label{sec:further}

\subsection{Examining a Distribution with a Deterministic Number of Weak Contacts}

\begin{figure}[!htbp]
    \centering
    \includegraphics[width=6.5in]{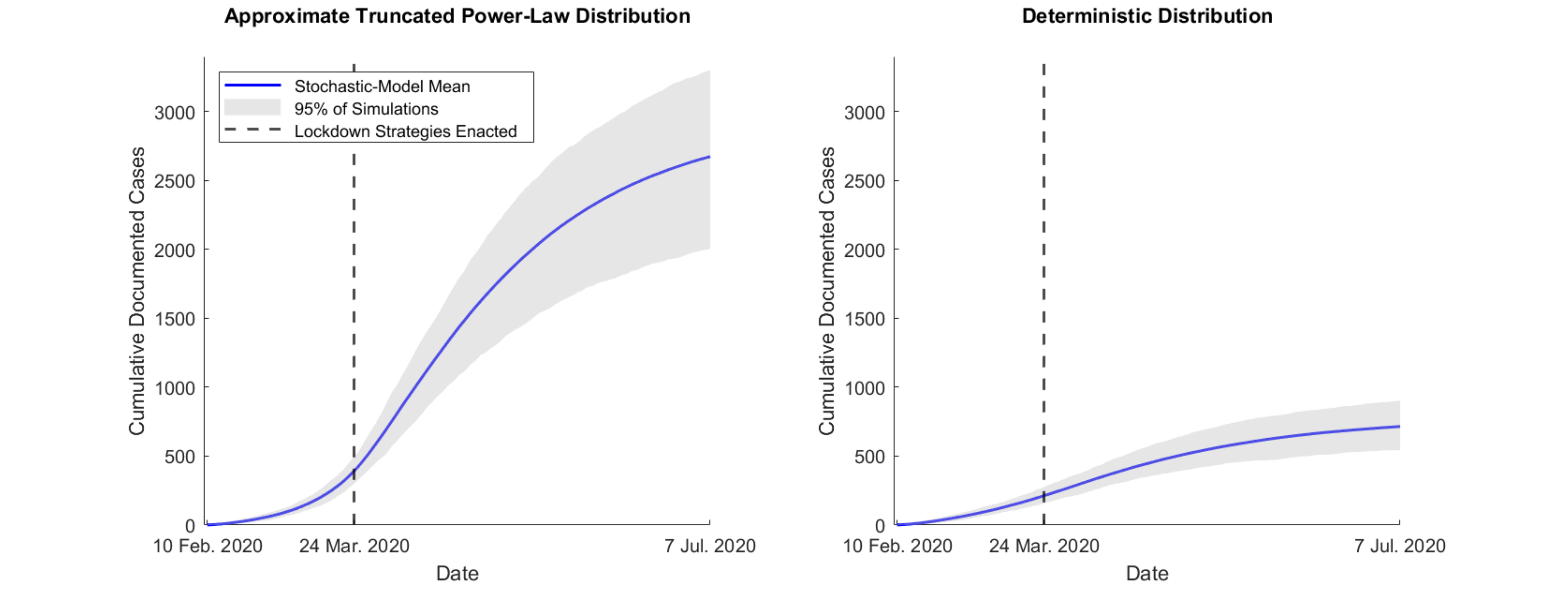}
    \caption{Comparison of a mean of 100 simulations when weak contacts are distributed according to an approximate truncated power-law distribution and a deterministic distribution. In both plots, the mean is depicted in blue and the gray window indicates the middle 95\% of these 100 simulations. On day 44 (24 March, 2020), all groups limit contacts and all individuals in caregiver--disabled interactions and interactions between essential workers and their weak contacts wear masks. 
	}
    \label{fig:fixed_dist}
    \end{figure}

The confidence window for the cumulative documented case counts is large. To determine the cause of this large variance, we run trials (see Fig.~\ref{fig:fixed_dist}) in which each subpopulation has a deterministic number of weak contacts that is equal to the mean values  in Section \ref{sec:modelSpecific}. When the weak-contact distribution is deterministic, we find that the variance in documented cases is much smaller than when weak contacts are distributed according to an approximate truncated power-law. Additionally, the using the deterministic distribution results in many fewer cases of the disease, which hardly spreads.

	%%%%
	
\subsection{Different Values of the Caregiver--Disabled Edge Weight $w_c$}

\begin{figure}[!htbp]
    \centering
    \includegraphics[width=7in]{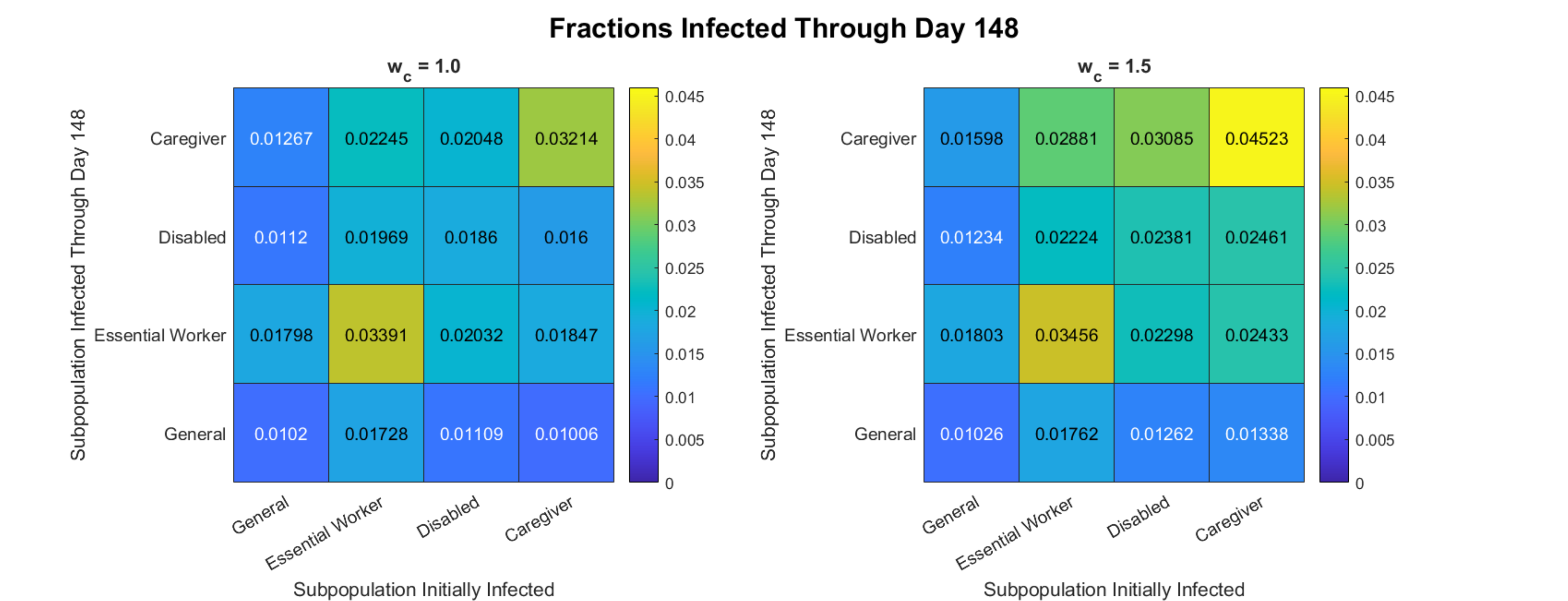}\vspace{2mm}
    \caption{Fraction of each subpopulation that is infected through day 148
    when all of the initially infected individuals {are in a single subpopulation} for (left) $w_c=1$ and (right) $w_c=1.5$. On day 44, all groups limit contacts and disabled people, caregivers, and essential workers wear masks.
    }
    \label{fig:care_weight_comparison}
    \end{figure}

The risk of COVID-19 infections in a caregiver--disabled interaction is larger than in an ordinary household interaction. In Fig.~\ref{fig:care_weight_comparison}, we compare our results for two different values of the caregiver--disabled edge weight $w_c$. The choice $w_c=1$ results in essential workers, who have many weak contacts, being the most potent disease spreaders among all subpopulations (except for spreading from caregivers to other caregivers). However, even the choice $w_c=1.5$ (which is smaller than the value $w_c = 2.27$ that we used in most of our computations) results in caregivers being the most {potent spreaders} of the disease to the disabled subpopulation.
	
%%%%
	
\end{document}